\RequirePackage{lineno}
\documentclass[twocolumn,aps,tightenlinefs,superscriptaddress,preprintnumbers]{revtex4}
\usepackage{graphicx,color,dcolumn,booktabs,bm}
\usepackage{longtable,lscape}
\usepackage{amsmath}
\usepackage{indentfirst}
\usepackage{epsfig}
\usepackage{feynmf}   
\usepackage{epstopdf}   
\usepackage{slashed}  
\usepackage{cases}
\usepackage{color}
\usepackage{multirow}
\usepackage{ulem}
\usepackage{verbatim}
\usepackage[colorlinks,linkcolor=red,anchorcolor=green,citecolor=blue]{hyperref}
\usepackage{mathrsfs}
\usepackage{rotating}
\usepackage{threeparttable}
\usepackage{lineno}
\usepackage{subfigure}
\usepackage{soul}
\usepackage{orcidlink}

\graphicspath{{ps}}

\newcommand{\lamc}{\Lambda_c^+}

\newcommand{\sgmeta}{\Sigma^+\eta}
\newcommand{\sgmetp}{\Sigma^+\eta'}
\newcommand{\sgmpiz}{\Sigma^+\pi^0}
\newcommand{\BF}{\mathcal{B}}

\begin{document}


\title{  \boldmath Measurements of branching fractions of $\Lambda_c^+ \to \Sigma^+ \eta$ and $\Lambda_c^+ \to \Sigma^+ \eta'$ and asymmetry parameters of $\Lambda_c^+ \to \Sigma^+ \pi^0$, $\Lambda_c^+ \to \Sigma^+ \eta$, and $\Lambda_c^+ \to \Sigma^+ \eta'$}

\noaffiliation
\author{S.~X.~Li\,\orcidlink{0000-0003-4669-1495}} 
\author{C.~P.~Shen\,\orcidlink{0000-0002-9012-4618}} 
\author{I.~Adachi\,\orcidlink{0000-0003-2287-0173}} 
\author{J.~K.~Ahn\,\orcidlink{0000-0002-5795-2243}} 
\author{H.~Aihara\,\orcidlink{0000-0002-1907-5964}} 
\author{D.~M.~Asner\,\orcidlink{0000-0002-1586-5790}} 
\author{H.~Atmacan\,\orcidlink{0000-0003-2435-501X}} 
\author{T.~Aushev\,\orcidlink{0000-0002-6347-7055}} 
\author{R.~Ayad\,\orcidlink{0000-0003-3466-9290}} 
\author{V.~Babu\,\orcidlink{0000-0003-0419-6912}} 
% ...
\collaboration{The Belle Collaboration}

\begin{abstract}
     We report a study of $\Lambda_c^+ \to \Sigma^+ \pi^0$, $\Lambda_c^+ \to \Sigma^+ \eta$, and $\Lambda_c^+ \to \Sigma^+ \eta'$ using the data sample corresponding to an integrated luminosity of 980 $\rm fb^{-1}$ collected with the Belle detector at the KEKB asymmetric-energy $e^+e^-$ collider. The branching fractions relative to $\Lambda_c^+ \to \Sigma^+ \pi^0$ are measured as:
     $\mathcal{B}_{\Lambda_c^+ \to \Sigma^+ \eta}/\mathcal{B}_{\Lambda_c^+ \to \Sigma^+ \pi^0}=0.25 \pm 0.03 \pm 0.01$
     and
     $\mathcal{B}_{\Lambda_c^+ \to \Sigma^+ \eta'}/\mathcal{B}_{\Lambda_c^+ \to \Sigma^+ \pi^0}=0.33 \pm 0.06 \pm 0.02$.
     Using $\mathcal{B}_{\Lambda_c^+ \to \Sigma^+ \pi^0}=(1.25 \pm 0.10)\%$, we obtain
     $\mathcal{B}_{\Lambda_c^+ \to \Sigma^+ \eta}=(3.14 \pm 0.35 \pm 0.17 \pm 0.25)\times10^{-3}$ and
     $\mathcal{B}_{\Lambda_c^+ \to \Sigma^+ \eta'}=(4.16 \pm 0.75 \pm 0.25 \pm 0.33)\times10^{-3}$.
     Here the uncertainties are statistical, systematic, and from $\mathcal{B}_{\Lambda_c^+ \to \Sigma^+ \pi^0}$, respectively. The ratio of the branching fraction of $\Lambda_c^+ \to \Sigma^+ \eta'$ with respect to that of $\Lambda_c^+ \to \Sigma^+ \eta$ is measured to be
     $\mathcal{B}_{\Lambda_c^+ \to \Sigma^+ \eta'}/\mathcal{B}_{\Lambda_c^+ \to \Sigma^+ \eta}=1.34 \pm 0.28 \pm 0.08$.
     We update the asymmetry parameter of $\Lambda_c^+ \to \Sigma^+ \pi^0$, $\alpha_{\Sigma^+ \pi^0} = -0.48 \pm 0.02 \pm 0.02$, with a considerably improved precision. The asymmetry parameters of $\Lambda_c^+ \to \Sigma^+ \eta$ and $\Lambda_c^+ \to \Sigma^+ \eta'$ are measured to be
     $\alpha_{\Sigma^+ \eta} = -0.99 \pm 0.03 \pm 0.05$
     and
     $\alpha_{\Sigma^+ \eta'} = -0.46 \pm 0.06 \pm 0.03$ for the first time.
\end{abstract}

\maketitle

\tighten

\section{\boldmath Introduction}
\begin{figure}[h!tbp]
           \includegraphics[width=4cm]{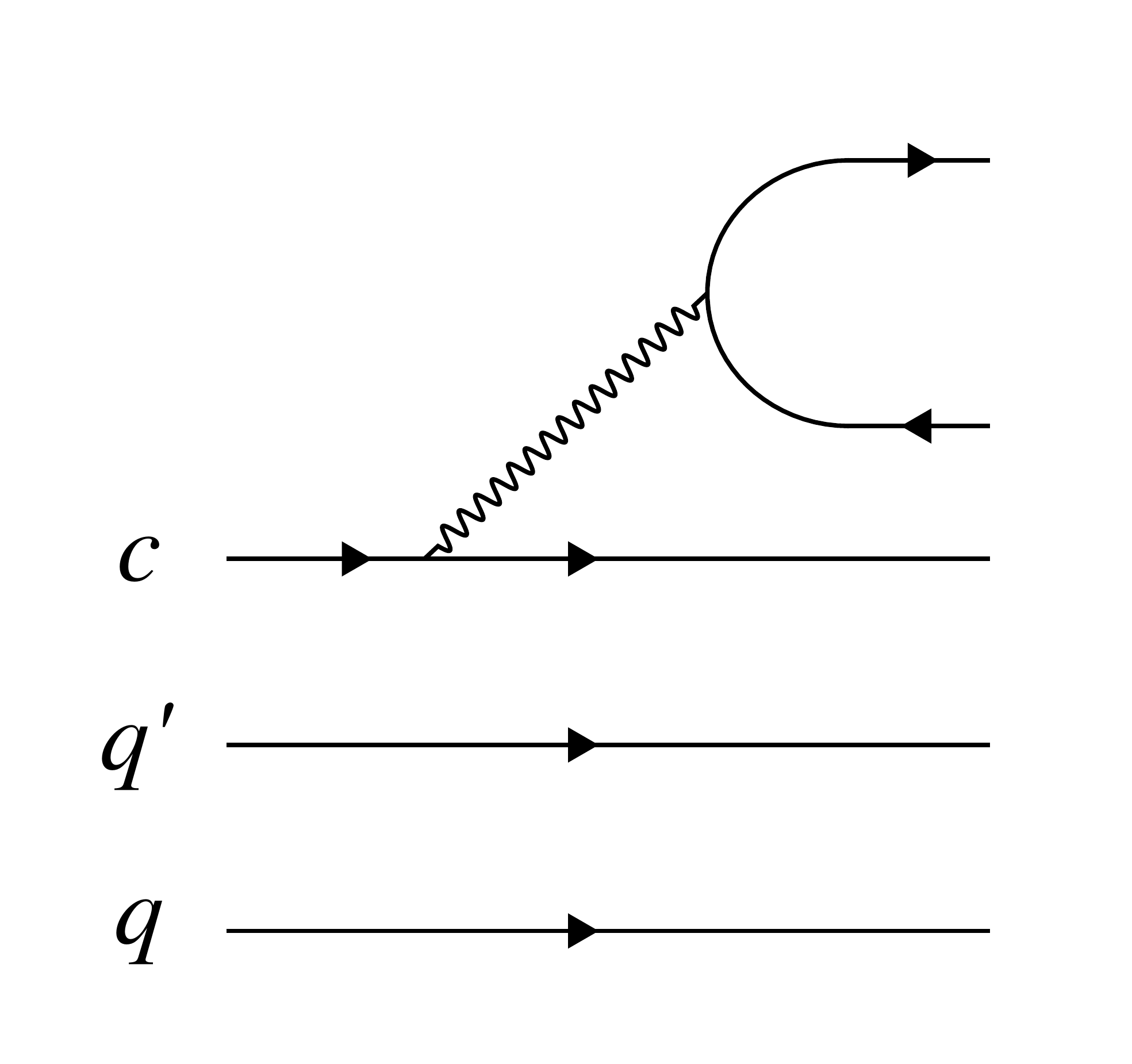}
           \includegraphics[width=4cm]{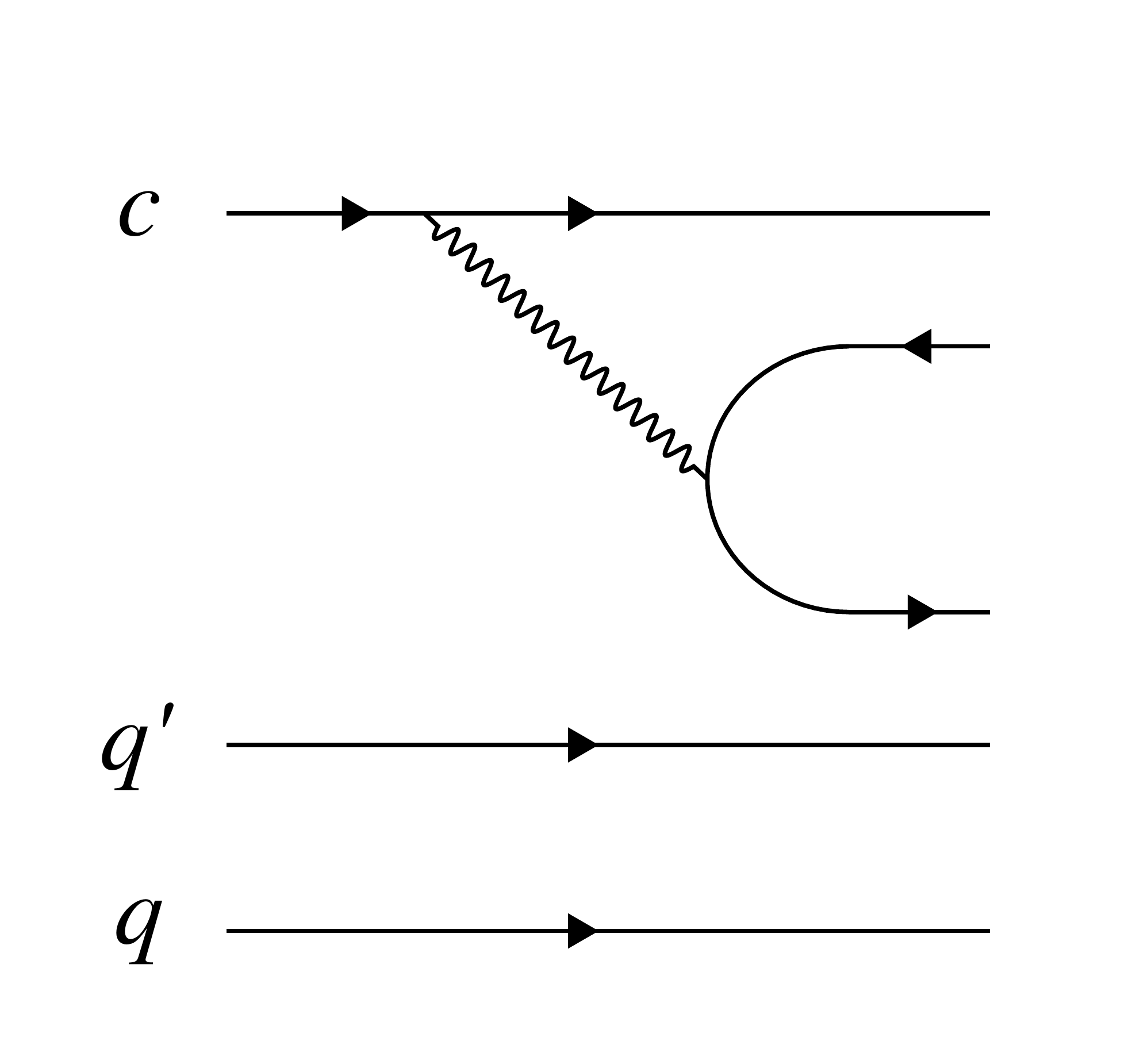}
           \put(-210,70){\bf (a)} \put(-90,70){\bf (b)}
           \vspace{0.0cm}

           \includegraphics[width=4cm]{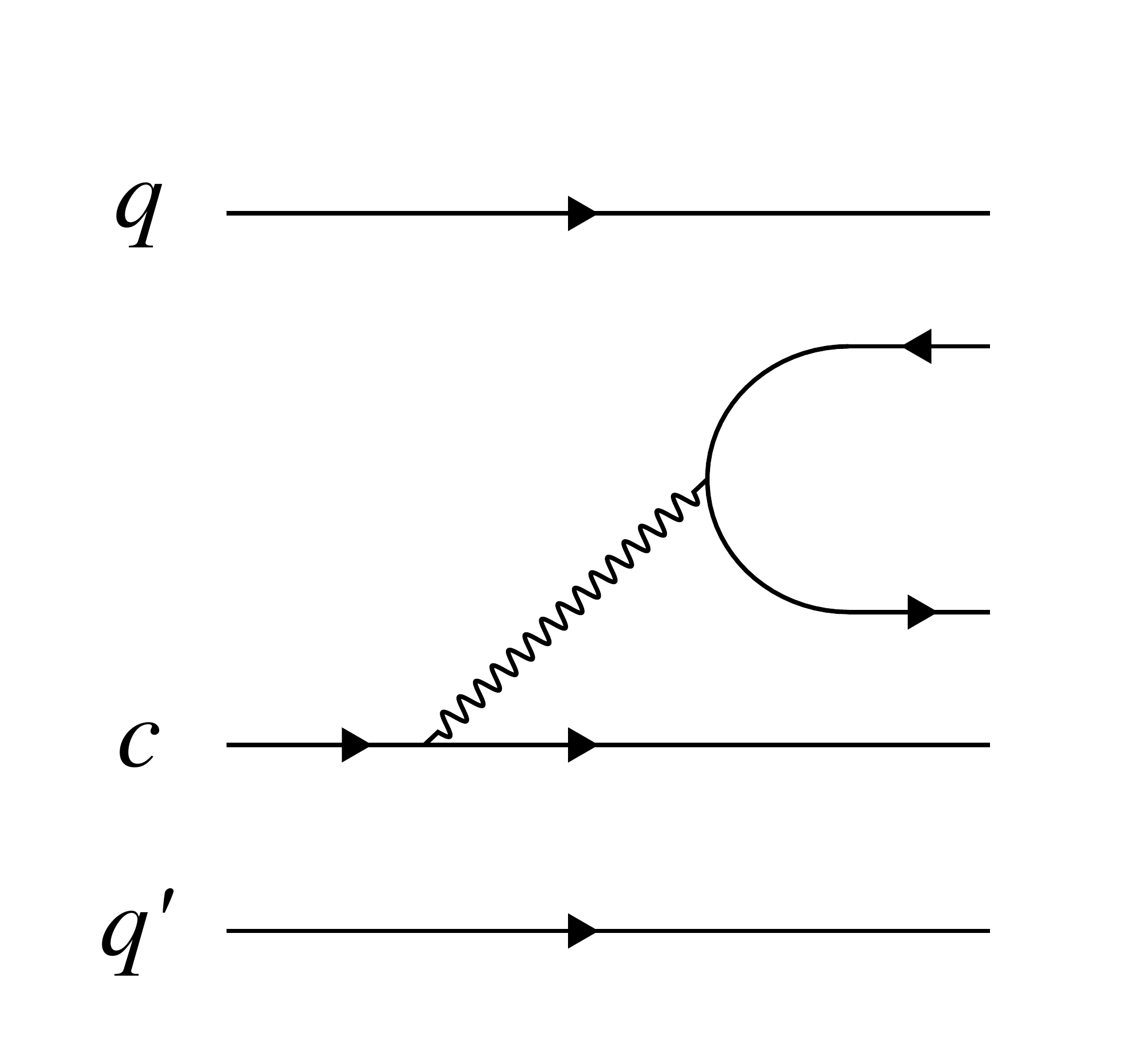}
           \includegraphics[width=4cm]{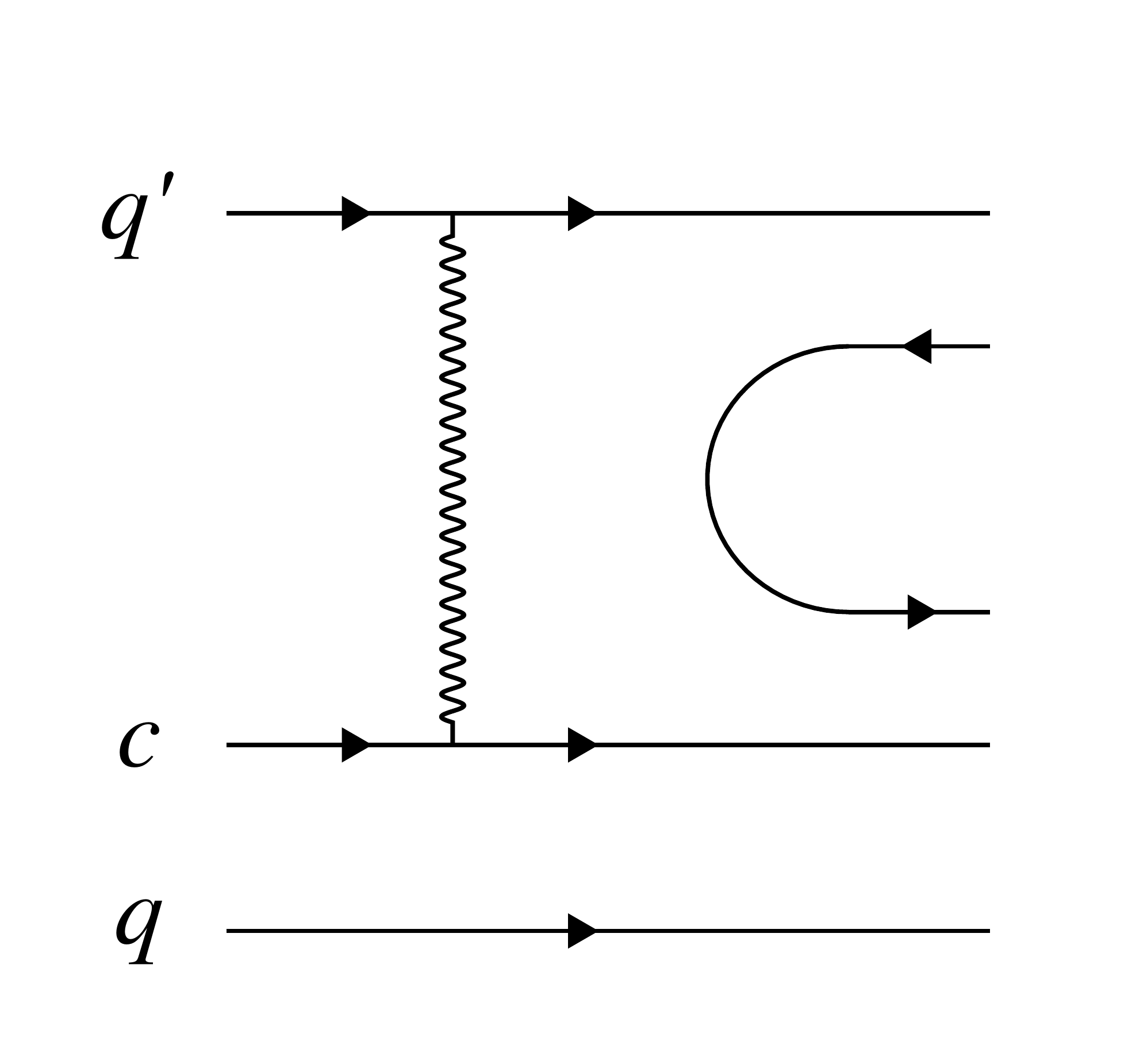}
           \put(-210,60){\bf (c)} \put(-90,60){\bf (d)}
           \vspace{0.0cm}

           \includegraphics[width=4cm]{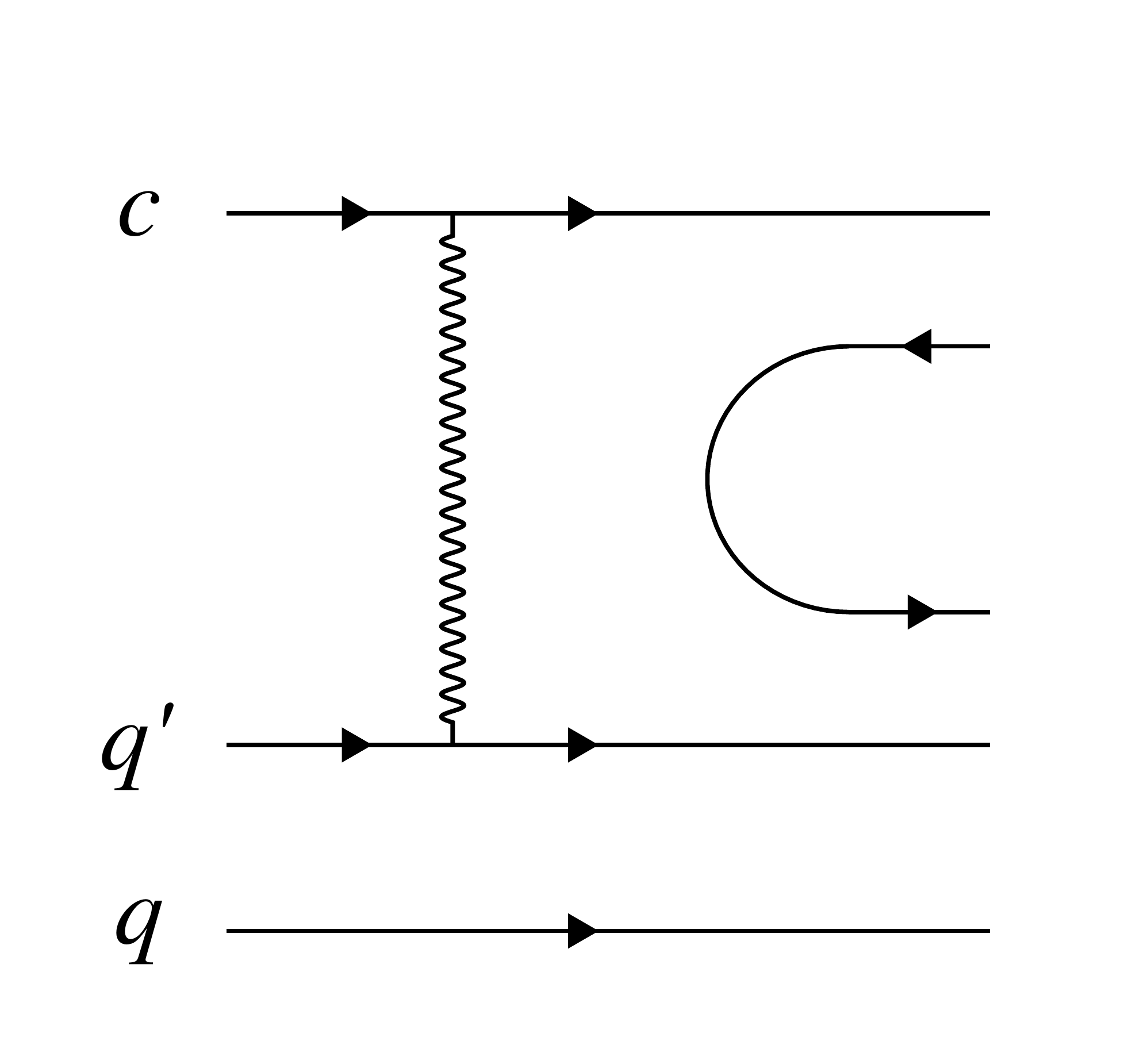}
           \includegraphics[width=4cm]{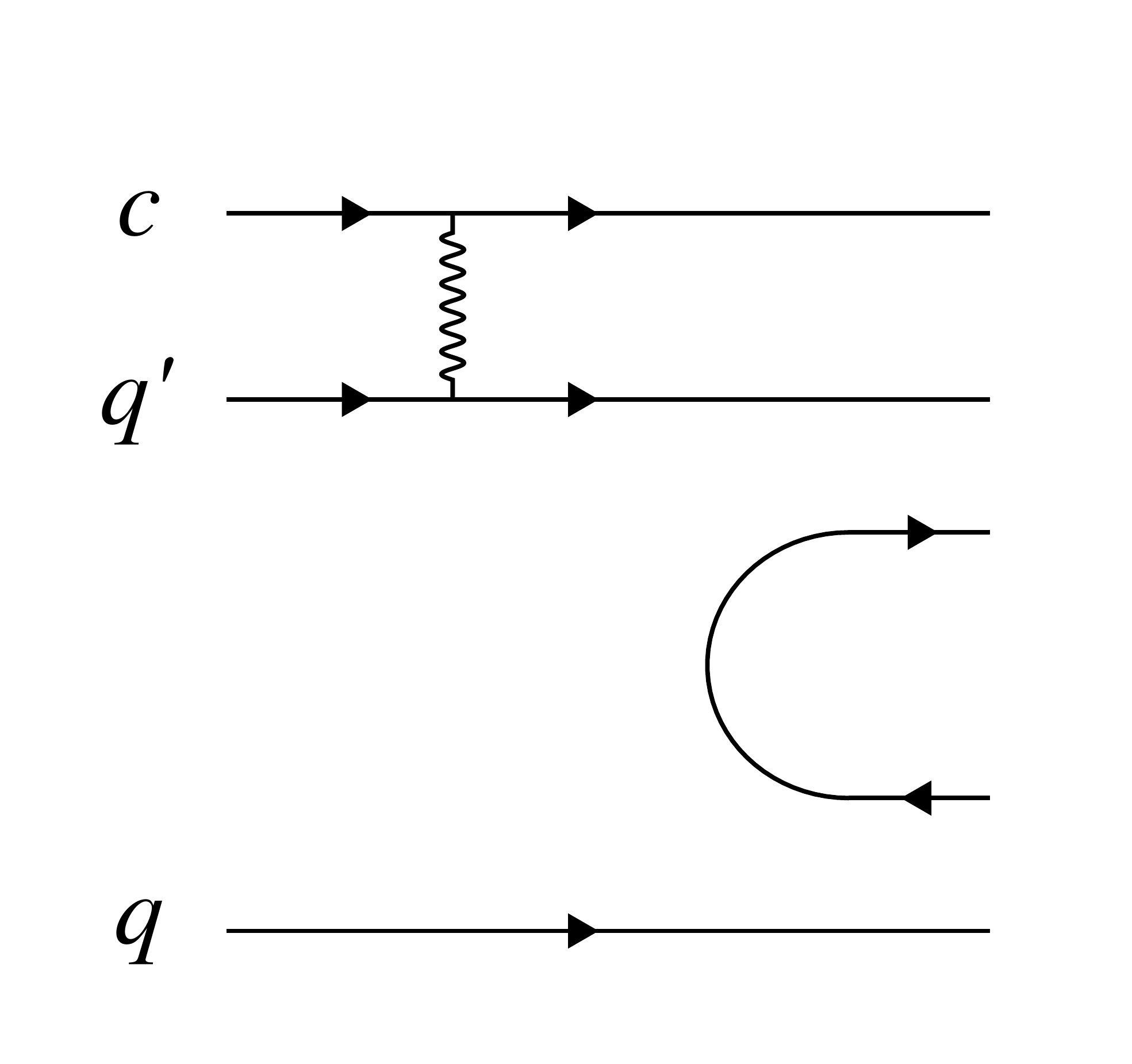}
           \put(-210,50){\bf (e)} \put(-90,50){\bf (f)}
           \caption{Topological diagrams contributing to $\mathbf{B}_c \to \mathbf{B} + \mathbf{M}$ decays: (a) external $W$ emission, (b) internal $W$ emission, (c) inner $W$ emission, and (d, e, f) $W$ exchange diagrams.}
           \label{fig:feynman}
     \end{figure}

\setstcolor{red}
     Hadronic weak decays of charmed baryons play an important role in understanding the decay dynamics of charmed baryons. For weak decays $\mathbf{B}_c \to \mathbf{B} + \mathbf{M}$, the decay mechanisms involved are $W$ emission and $W$ exchange, where $\mathbf{B}_c$, $\mathbf{B}$, and $\mathbf{M}$ represent the charmed baryon, baryon, and pseudoscalar or vector meson, respectively. The topological diagrams are shown in Fig.~\ref{fig:feynman}. Among them, external $W$ emission (Fig.~\ref{fig:feynman} (a)) and internal $W$ emission (Fig.~\ref{fig:feynman} (b)) are factorizable, while inner $W$ emission (Fig.~\ref{fig:feynman} (c)) and $W$ exchange (Fig.~\ref{fig:feynman} (d, e, f)) give nonfactorizable contributions. For Cabibbo-allowed two-body decays $\Lambda_c^+ \to \Sigma^+ \eta$ and $\Lambda_c^+ \to \Sigma^+ \eta'$, all the nonfactorizable diagrams contribute to these two decays.

     To describe the nonfactorizable effects in hadronic decays of charmed baryons, various approaches have been developed, including the covariant confined quark model~\cite{theo1,theo2}, the pole model~\cite{theo3,theo4,theo5,theo6,theo7,theo8}, current algebra~\cite{theo9,theo10} and the SU(3) flavor symmetry~\cite{theo11}. Theoretical calculations on branching fractions of the above two signal decays based on various approaches are listed in Table~\ref{tab:bfalpha:theory:experiment}. The two branching fractions were measured to be $\mathcal{B}_{\Lambda_c^+ \to \Sigma^+ \eta}=(0.44\pm0.20)\%$ and $\mathcal{B}_{\Lambda_c^+ \to \Sigma^+ \eta'}=(1.5\pm0.6)\%$, with poor precision~\cite{cpcResult}. Their ratio is $\mathcal{B}_{\Lambda_c^+ \to \Sigma^+ \eta'}/\mathcal{B}_{\Lambda_c^+ \to \Sigma^+ \eta}=3.5\pm2.1\pm0.4$~\cite{ratiocite}.

     The decay asymmetry parameter $\alpha$ was introduced by Lee and Yang to study the parity-violating and parity-conserving amplitudes in weak hyperon decays~\cite{alphadefine1}. Charge conjugation and parity ($CP$) symmetry implies that the baryon decay asymmetry parameter $\alpha$ equals that of the antibaryon $\bar{\alpha}$ but with an opposite sign~\cite{nature}. Measuring the asymmetry parameter reveals the degree of parity violation in baryon decays and enables tests of $CP$ symmetry used to search for physics beyond the standard model.

     The weak decay asymmetry parameter $\alpha_{\Sigma^+ h^0}$ of the $\Lambda_c^+ \to \Sigma^+h^0$, $\Sigma^+ \to p\pi^0$ ($h^0=\pi^0$, $\eta$, or $\eta'$) decays determines the differential decay rate~\cite{alphacite},
          \begin{equation}
                \frac{dN}{d\cos\theta_{\Sigma^+}} \propto 1 + \alpha_{\Sigma^+h^0}\alpha_{p\pi^0}\cos\theta_{\Sigma^+},
          \end{equation}
     where $\theta_{\Sigma^+}$ is the angle between the proton momentum vector and the opposite of the $\Lambda_c^+$ momentum vector in the $\Sigma^+$ rest frame~\cite{expalpha} and $\alpha_{p\pi^0}=-0.983 \pm 0.013$ is the averaged asymmetry parameter of $\Sigma^+ \to p\pi^0$ and its charge-conjugated mode~\cite{pdg}.
     It is related to the parity-violating amplitude $A_1$ and parity-conserving amplitude $A_2$ in the decay through~\cite{alphacite}
     \begin{equation}
           \alpha_j = 2\frac{\kappa \rm Re(\it A_1A_2^*)}{|A_1|^2+\kappa^2|A_2|^2},
     \end{equation}
     where $\kappa = pc/(E+mc^2)$ with $p$, $E$, and $m$ being the momentum, energy, and mass of the daughter baryon in the rest frame of the parent particle and $c$ is the speed of light, and $j$ denotes the particular decay mode.
     The asymmetry parameter of $\Lambda_c^+ \to \Sigma^+ \pi^0$ has been measured to be $-0.55\pm0.11$, with an uncertainty of about 20\%~\cite{pdg}. Up to now, little experimental information on measurements of asymmetry parameters of $\Lambda_c^+ \to \Sigma^+ \eta$ and $\Lambda_c^+ \to \Sigma^+ \eta'$ is available. Theoretical predictions on the asymmetry parameters are calculated based on various models~\cite{theo1,theo2,theo7,theo8,theo10,theo11}, as listed in Table~\ref{tab:bfalpha:theory:experiment}.

     \begin{table*}[htbp]
             \caption{\label{tab:bfalpha:theory:experiment}
              Branching fractions (upper entry) in \%  and asymmetry parameters (lower entry) in various approaches and in experiment.}
             \setlength{\tabcolsep}{2.8mm}{
             \begin{tabular}{r@{$\to$}l r@{.}l r@{.}l r@{.}l r@{.}l r@{.}l r@{$\pm$}l r@{$\pm$}l}\hline \hline
              \multicolumn{2}{c}{Decay} &\multicolumn{2}{c}{K$\rm \ddot{o}$rner~\cite{theo1}} &\multicolumn{2}{c}{Ivanov~\cite{theo2}} &\multicolumn{2}{c}{$\rm \ddot{Z}$enczykowski~\cite{theo7}} &\multicolumn{2}{c}{Sharma~\cite{theo8}} &\multicolumn{2}{c}{Zou~\cite{theo10}} &\multicolumn{2}{c}{Geng~\cite{theo11}} &\multicolumn{2}{c}{Experiment~\cite{pdg}}\\ \hline
              $\Lambda_c^+$ & $\Sigma^+ \eta$  &0&16  &0&11  &0&90  &0&57 &0&74 &0.32&0.13 &0.44&0.20 \\
              $\Lambda_c^+$ & $\Sigma^+ \eta'$ &1&28  &0&12  &0&11  &0&10 &\multicolumn{2}{c}{--} &1.44&0.56 &1.5&0.6 \\ \hline
              $\Lambda_c^+$ & $\Sigma^+ \pi^0$  &0&70 &0&43  &0&39  &$-0$&31 &$-0$&76 &$-0.35$&0.27 &$-0.55$&0.11 \\
              $\Lambda_c^+$ & $\Sigma^+ \eta$  &0&33  &0&55  &0&00  &$-0$&91 &$-0$&95 &$-0.40$&0.47 &\multicolumn{2}{c}{--} \\
              $\Lambda_c^+$ & $\Sigma^+ \eta'$ &$-0$&45 &$-0$&05 &$-0$&91 &0&78 &0&68 &\multicolumn{2}{c}{$1.00^{+0.00}_{-0.17}$} &\multicolumn{2}{c}{--} \\ \hline \hline
              \end{tabular}
              }
      \end{table*}

      The isospin symmetry demands that the branching fractions and the asymmetry parameters of $\Lambda_c^+ \to \Sigma^+ \pi^0$ and $\Lambda_c^+ \to \Sigma^0 \pi^+$ shall be equal~\cite{theo11}. Any significant discrepancy would be an indication of isospin breaking, which could be caused by the mass difference between up and down quarks, the electromagnetic interaction, or new physics. Precise measurements of the branching fractions and asymmetry parameters of $\Lambda_c^+ \to \Sigma^+ \pi^0$ and $\Lambda_c^+ \to \Sigma^0 \pi^+$ will provide an excellent test of the isospin symmetry.

     In this study, we measure the branching fractions of $\Lambda_c^+ \to \Sigma^+ \eta$ and $\Lambda_c^+ \to \Sigma^+ \eta'$ relative to the $\Lambda_c^+ \to \Sigma^+ \pi^0$ based on the full Belle data. We update the measurement of the asymmetry parameter of $\Lambda_c^+ \to \Sigma^+ \pi^0$ with higher statistics, and measure the asymmetry parameters of $\Lambda_c^+ \to \Sigma^+ \eta$ and $\Lambda_c^+ \to \Sigma^+ \eta'$ for the first time.

\section{\boldmath The data sample and the belle detector}

     This analysis is based on a data sample corresponding to an integrated luminosity of 980 $\rm fb^{-1}$, collected with the Belle detector at the KEKB asymmetric-energy $e^+e^-$ collider~\cite{KEKB1,KEKB2}. About 70\% of the data were recorded at the $\Upsilon(4S)$ resonance, and the rest were collected at other $\Upsilon(nS)$ ($n$ = 1,~2,~3, or 5) states or at center-of-mass (c.m.) energies a few tens of MeV below the $\Upsilon(4S)$ or the $\Upsilon(nS)$ peaks.

     The Belle detector is a large-solid-angle magnetic spectrometer that consists of a silicon vertex detector (SVD), a 50-layer central drift chamber (CDC), an array of aerogel threshold Cherenkov counters (ACC), a barrel-like arrangement of time-of-flight scintillation counters (TOF), and an electromagnetic calorimeter comprised of CsI(Tl) crystals (ECL) located inside a superconducting solenoid coil that provides a 1.5~T magnetic field. An iron flux-return located outside of the coil is instrumented to detect $K_L^0$ mesons and to identify muons. The detector is described in detail elsewhere~\cite{Belle1,Belle2}.
     The position of the nominal interaction point, where electrons and positrons collide, is defined as the origin of the coordinate system, and the axis aligning with the direction opposite the $e^+$ beam is defined as the $z$ axis. The ECL is divided into three regions spanning $\theta$, the angle of inclination in the laboratory frame with respect to the $z$ axis, with the backward endcap, barrel and forward endcap regions covering the ranges of $-0.91 < \cos\theta < -0.63$, $-0.63 < \cos\theta < 0.85$, and $0.85 < \cos\theta < 0.98$, respectively.

     Monte Carlo (MC) simulated events are used to optimize the selection criteria, study background contributions, and determine the signal reconstruction efficiency. Samples of simulated signal MC events are generated by {\sc EvtGen}~\cite{evtgen} and propagated through a detector simulation based on {\sc geant3}~\cite{geant3}. The $e^+e^- \to c\bar{c}$ events are simulated using {\sc pythia}~\cite{pythia}; the decays $\eta' \to \eta\pi^+\pi^-$ and $\pi^0/\eta \to \gamma\gamma$ are generated with a phase space model; the decays $\Lambda_c^+ \to \Sigma^+ h^0$ and $\Sigma^+ \to p\pi^0$ are first generated with a phase space model, afterwards corrected using the measured asymmetry parameter to ensure a correct angular distribution. We take into account the effect of final-state radiation from charged particles by using the {\sc photos} package~\cite{photons}. Simulated samples of $\Upsilon(4S)\to B^{+}B^{-}/B^{0}\bar{B}^{0}$, $\Upsilon(5S) \to B_{s}^{(*)}\bar{B}_{s}^{(*)}/B^{(*)}\bar{B}^{(*)}(\pi)/\Upsilon(4S)\gamma$, $e^+e^- \to q\bar{q}$ $(q=u,~d,~s,~c)$ at $\sqrt{s}$ = 10.52, 10.58, and 10.867~GeV, and $\Upsilon(1S,~2S,~3S)$ decays, corresponding to four times the integrated luminosity of each dataset and normalized to the same integrated luminosity as data, are used to develop the selection criteria and perform the background study. Charge-conjugate modes are implied throughout this paper unless otherwise stated.

\section{\boldmath Event selection}

     We reconstruct the decays $\Lambda_c^+ \to \Sigma^+ \pi^0$, $\Sigma^+ \eta$, and $\Sigma^+ \eta'$, with $\Sigma^+ \to p\pi^0$, $\eta' \to \eta\pi^+\pi^-$, and $\pi^0/\eta \to \gamma\gamma$.

     A set of event selection criteria is chosen to enhance the $\Lambda_c^+$ signal while reducing combinatorial background events. These criteria are determined by maximizing the figure-of-merit $\epsilon/(\frac{5}{2}+\sqrt{n_{\rm B}})$~\cite{fomcite} for each variable under consideration in an iterative fashion, where $\epsilon$ is the signal efficiency; $n_{\rm B}$ is the number of background events expected in the $\Lambda_c^+$ signal region ($\pm 2.5\sigma$ from $\Lambda_c^+$ nominal mass~\cite{pdg}).

     Charged tracks are identified as proton or pion candidates using information from the tracking (SVD, CDC) and charged-hadron identification (ACC, TOF, CDC) systems combined into a likelihood, $\mathcal{L}(h:h') = \mathcal{L}(h)/(\mathcal{L}(h)+\mathcal{L}(h'))$ where $h$ and $h'$ are $\pi$, $K$, or $p$ as appropriate~\cite{pidcode}. Tracks having $\mathcal{L}(p:\pi)>0.9$ and $\mathcal{L}(p:K)>0.9$ are identified as proton candidates and having $\mathcal{L}(\pi:p) > 0.4$ and $\mathcal{L}(\pi:K) > 0.4$ are identified as pion candidates. A likelihood ratio for electron identification, $\mathcal{R}(e)$ formed from ACC, CDC, and ECL information~\cite{eidcode}, is required to be less than 0.9 for all charged tracks to suppress electrons. The identification efficiencies of $p$ and $\pi$ are 80\% and 95\%, respectively. The probabilities of misidentifying $h$ as $h'$, $P(h\to h')$, are estimated to be 2\% for $P(p\to \pi)$, 6\% for $P(K\to \pi)$, 1\% for $P(K\to p)$, and 0.4\% for $P(\pi\to p)$. For each charged track, the number of SVD hits in both the $z$ direction and the transverse $x-y$ plane is required at least one. For pion candidates, we require the distance of the closest approach with respect to the interaction point (IP) along the $z$ axis and in the $x-y$ plane to be less than 2.0 cm and 0.1 cm, respectively.

     Photon candidates are selected from ECL clusters not matched to a CDC track trajectory. To reject neutral hadrons, the ratio of the energy deposited in the central $3\times3$ array of ECL crystals to the total energy deposited in the enclosing $5\times5$ array of crystals is required to be at least 0.8 for each photon candidate.

     The $\pi^0$ candidates are reconstructed by combining two photons with an invariant-mass $M(\gamma\gamma)$ between 0.124 and 0.145 GeV/$c^2$ (around $\pm 2.5\sigma$ of the nominal $\pi^0$ mass~\cite{pdg}). We perform a mass-constrained fit to each $\pi^0$ candidate to improve the momentum resolution.
     The daughter photons of $\pi^0$ originating from $\Sigma^+$ decay are required to have an energy greater than 50 MeV, while for those from $\Lambda_c^+$ decay, the energies in the barrel and endcaps are required to exceed 50 and 150 MeV, respectively. The reconstructed $\pi^0$ momentum in the c.m. frame must exceed 0.1 and 0.8 GeV/$c$ for candidates from $\Sigma^+$ and $\Lambda_c^+$ decays, respectively.

     The $\eta$ candidates are reconstructed by combining two photons with an invariant-mass $M(\gamma\gamma)$ between 0.523 and 0.567 GeV/$c^2$ (around $\pm 2.5\sigma$ of the nominal $\eta$ mass~\cite{pdg}). To improve the momentum resolution, a mass-constrained fit is performed for $\eta$ candidates. We require the energy of daughter photons of $\eta$ originating from $\eta'\to \eta\pi^+\pi^-$ and $\Lambda_c^+\to \Sigma^+\eta$ decays to be greater than 210 and 260 MeV, respectively. The reconstructed $\eta$ momentum in the c.m. frame is required to be  greater than 0.8 GeV/$c$ for $\Lambda_c^+ \to \Sigma^+\eta$ decay. To further suppress background photons from $\pi^0$ decay, we exclude any photon that, in combination with another photon in the event, has a probability of the reconstructed particle to be $\pi^0$-like greater than 0.6. The probability is calculated using the invariant-mass of the photon pair, the energy of the photon in the laboratory frame, and the angle with respect to the beam direction in the laboratory frame~\cite{pi0veto}.

     The $\eta'$ candidates are reconstructed by combining two oppositely charged $\pi$ candidates with an $\eta$ candidate. The candidates with an invariant-mass $M(\eta\pi^+\pi^-)$ within (0.953, 0.964) GeV/$c^2$ are retained (around $\pm 2.5\sigma$ of nominal $\eta'$ mass~\cite{pdg}).

     The $\Sigma^+$ candidates are formed by combining a $\pi^0$ candidate with a proton, with $M(p\pi^0)$ lying between 1.179 and 1.197 GeV/$c^2$ (around $\pm 2.5\sigma$ of nominal $\Sigma^+$ mass~\cite{pdg}).  The $\Sigma^+ \to p\pi^0$ reconstruction relies on the long lifetime of the hyperon. We require the proton candidates to have a large ($>1$ mm) distance of closest approach to the IP. The $\Sigma^+$ trajectory is approximated by a straight line from the IP in the direction of the reconstructed $\Sigma^+$ three-momentum and intersected with the proton path. This point is taken as an estimate of the $\Sigma^+$ decay vertex and the $\pi^0$ is re-fit assuming that the $\gamma\gamma$ pair originates from this vertex rather than from the IP. Only $\Sigma^+$ candidates with a positive flight length from the IP to the decay vertex are retained. We require a minimum $\Sigma^+$ momentum of 1.2 GeV/$c$ in the c.m. frame.

     The $\Lambda_c^+$ candidates are reconstructed by combining a $\Sigma^+$ candidate with a $\pi^0$, $\eta$, or $\eta'$ candidate. To suppress background especially from $B$-meson decays, we require the scaled momentum $x_p > 0.5$, defined as $x_{p} = p^*/p_{\rm max}$, where $p^*$ is the three-momentum of $\Lambda_c^+$ in the c.m. frame; and $p_{\rm max} = \sqrt{(E_{\rm beam}/c)^2-(Mc)^2}$ is its maximum momentum calculated by the beam energy ($E_{\rm beam}$) in the c.m. frame and the invariant mass ($M$) of $\Sigma^+h^0$ candidates. With the requirement on $x_p$, the background level can be suppressed by around 90\% based on the study of inclusive MC samples. After applying all selections, the fractions of the number of events having more than one $\Lambda_c^+$ candidate to those surviving the selections for $\Sigma^+ \pi^0$, $\Sigma^+ \eta$, and $\Sigma^+ \eta'$ modes are 0.8\%, 2.2\%, and 5.3\% , respectively. All the $\Lambda_c^+$ candidates in the mass region of (2.10, 2.45) GeV/$c^2$ are retained for further analysis.

\section{\boldmath Background study}
\label{sec:background}
\begin{figure*}[htbp]
           \begin{center}
              \includegraphics[width=5.9cm]{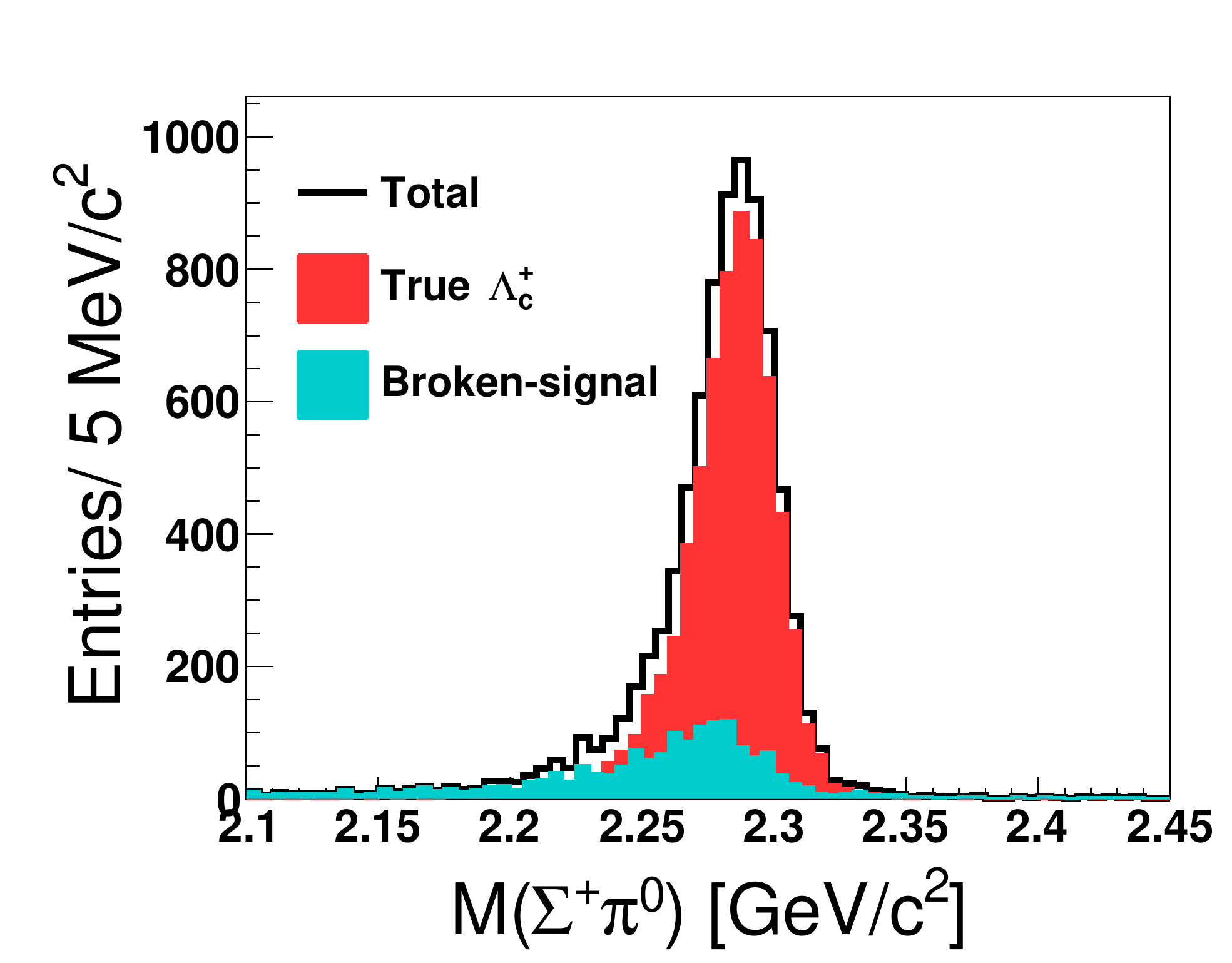}
              \includegraphics[width=5.9cm]{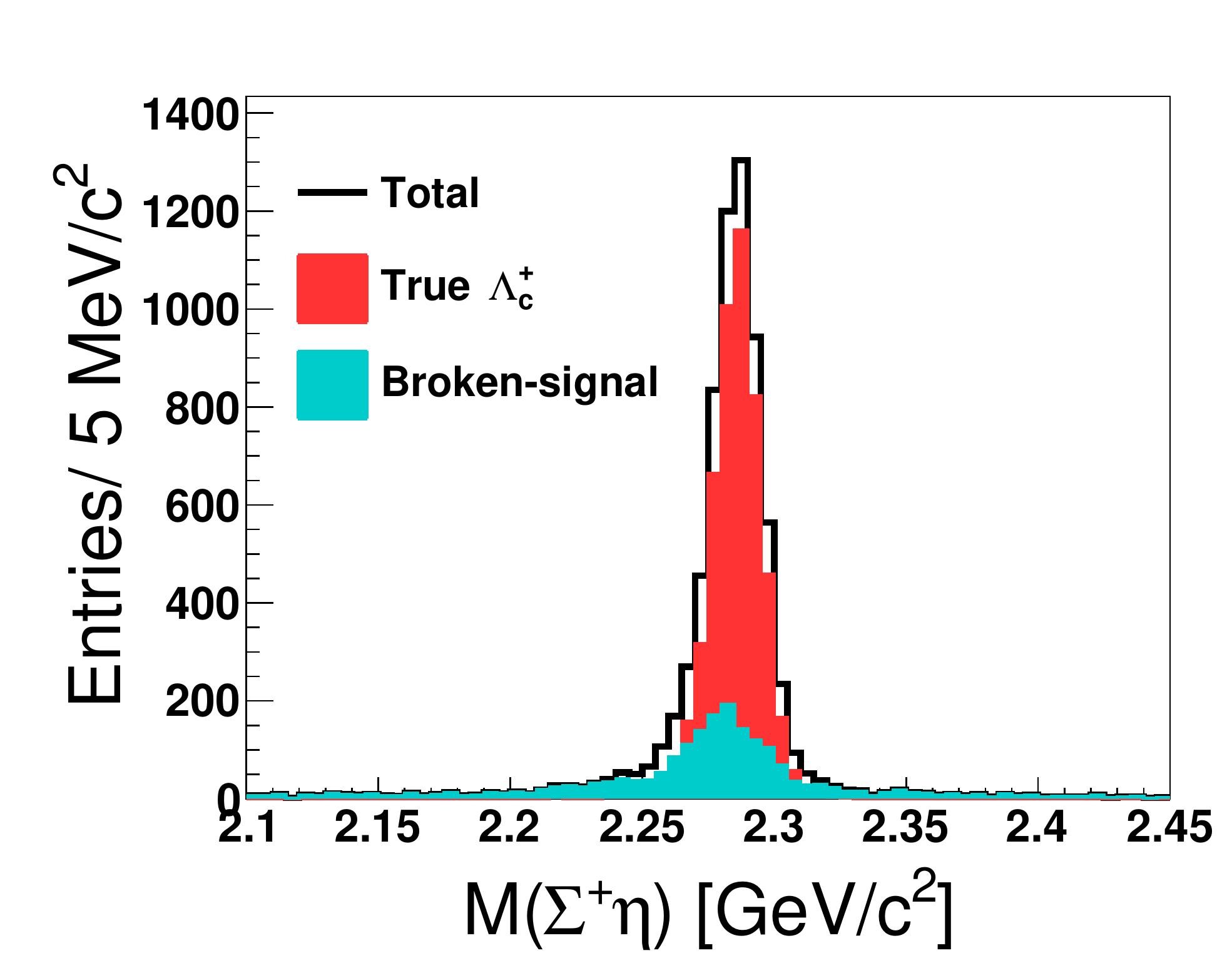}
              \includegraphics[width=5.9cm]{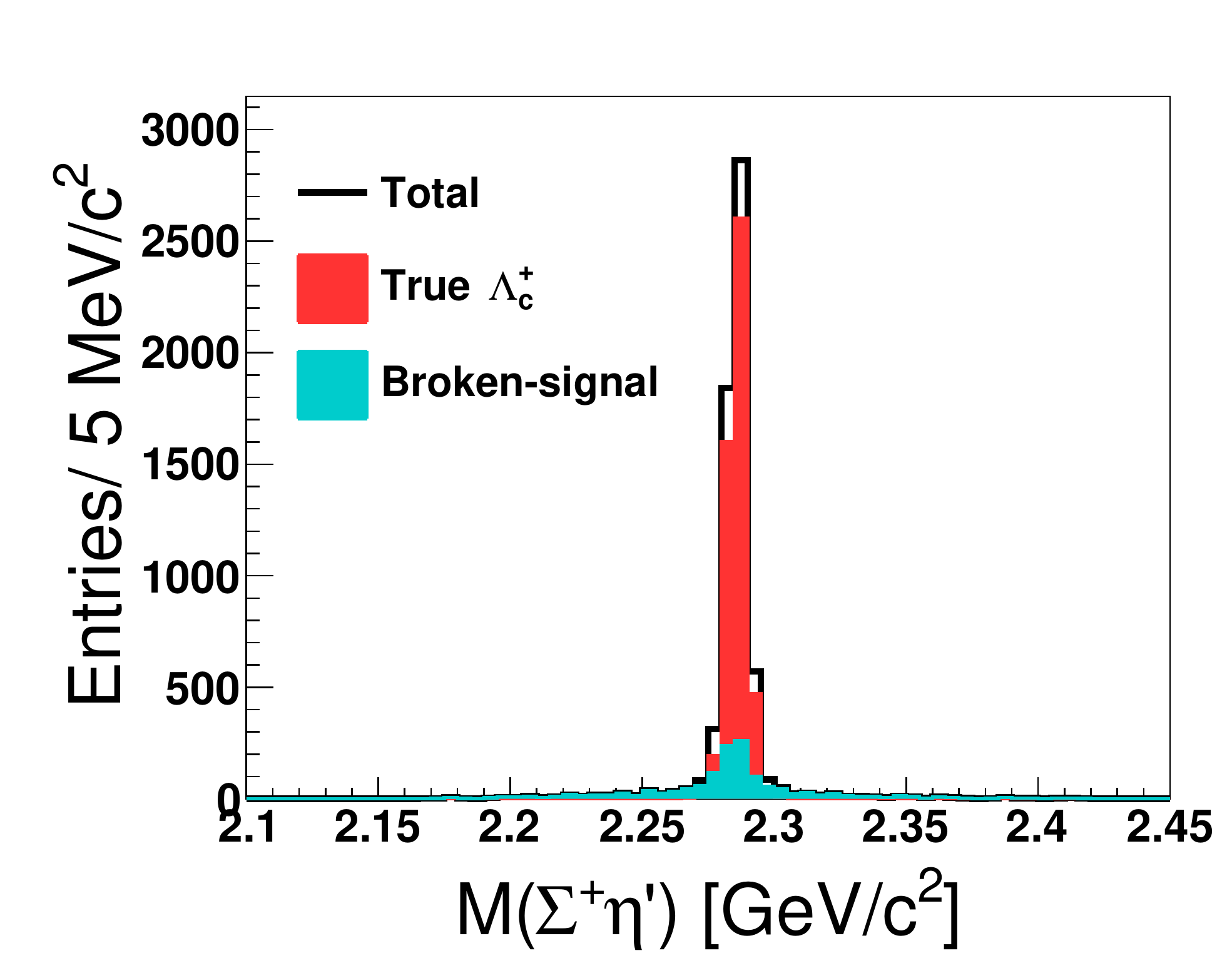}
              \put(-380,90){\large \bf (a)} \put(-210,90){\large \bf (b)}  \put(-40,90){\large \bf (c)}
              \caption{Distributions of $\Lambda_c^+$ invariant mass from signal MC simulation for (a) $\Lambda_c^+ \to \Sigma^+ \pi^0$, (b) $\Lambda_c^+ \to \Sigma^+ \eta$, and (c) $\Lambda_c^+ \to \Sigma^+ \eta'$ modes. The red histogram shows the true $\Lambda_c^+$ shape after applying MC-truth matching and the dark cyan histogram shows the shape of broken-signal contribution.}
              \label{fig:true:wrong}
           \end{center}
     \end{figure*}

     Using simulations, we search for possible background components that might peak in the $\Lambda_c^+$ mass region.
     A peaking background denoted as the broken-signal contribution, mainly caused by the wrong combination of a real photon with a fake photon to form a $\pi^0$ candidate from the $\Sigma^+$ decay, is observed in signal MC simulation. The distributions of the true $\Lambda_c^+$ events and the broken-signal events for the three modes are shown in Fig.~\ref{fig:true:wrong}, where the red histogram shows the true $\Lambda_c^+$ shape and the dark cyan histogram shows the shape of broken-signal contribution. The resolution of the broken-signal is obviously wider than that of the true signal, implying we can separate it from the true signal. The shape of the broken-signal contribution is extracted from MC simulation directly and is treated as a separate background component in the final $\Lambda_c^+$ signal yield extraction. The $M(\Sigma^+h^0)$ distributions of the MC sample and data in two-dimensional sidebands described below are consistent, suggesting that the MC simulation is reliable for this broken-signal contribution.

\begin{figure*}[htbp]
           \begin{center}
              \includegraphics[width=5.9cm]{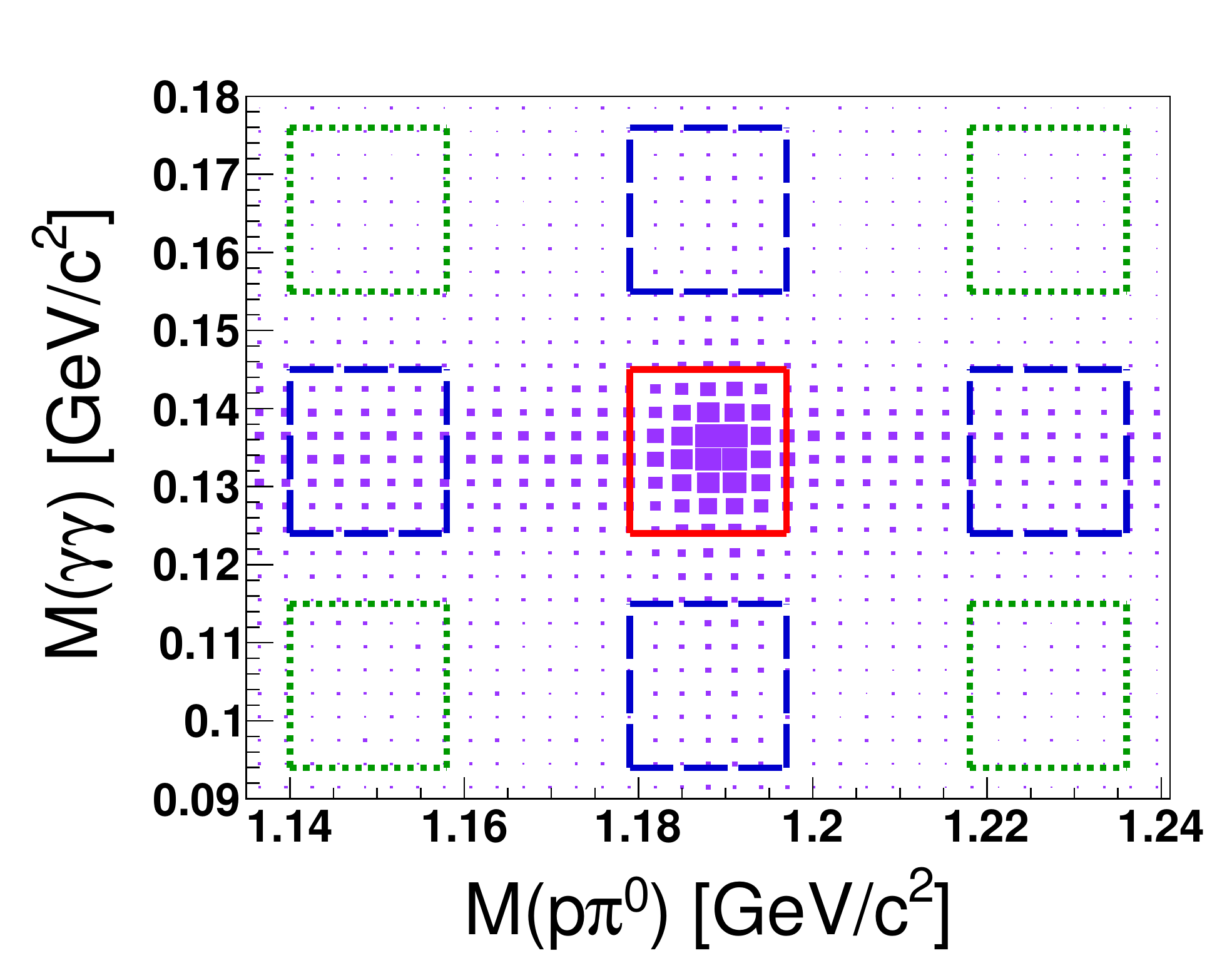}
              \includegraphics[width=5.9cm]{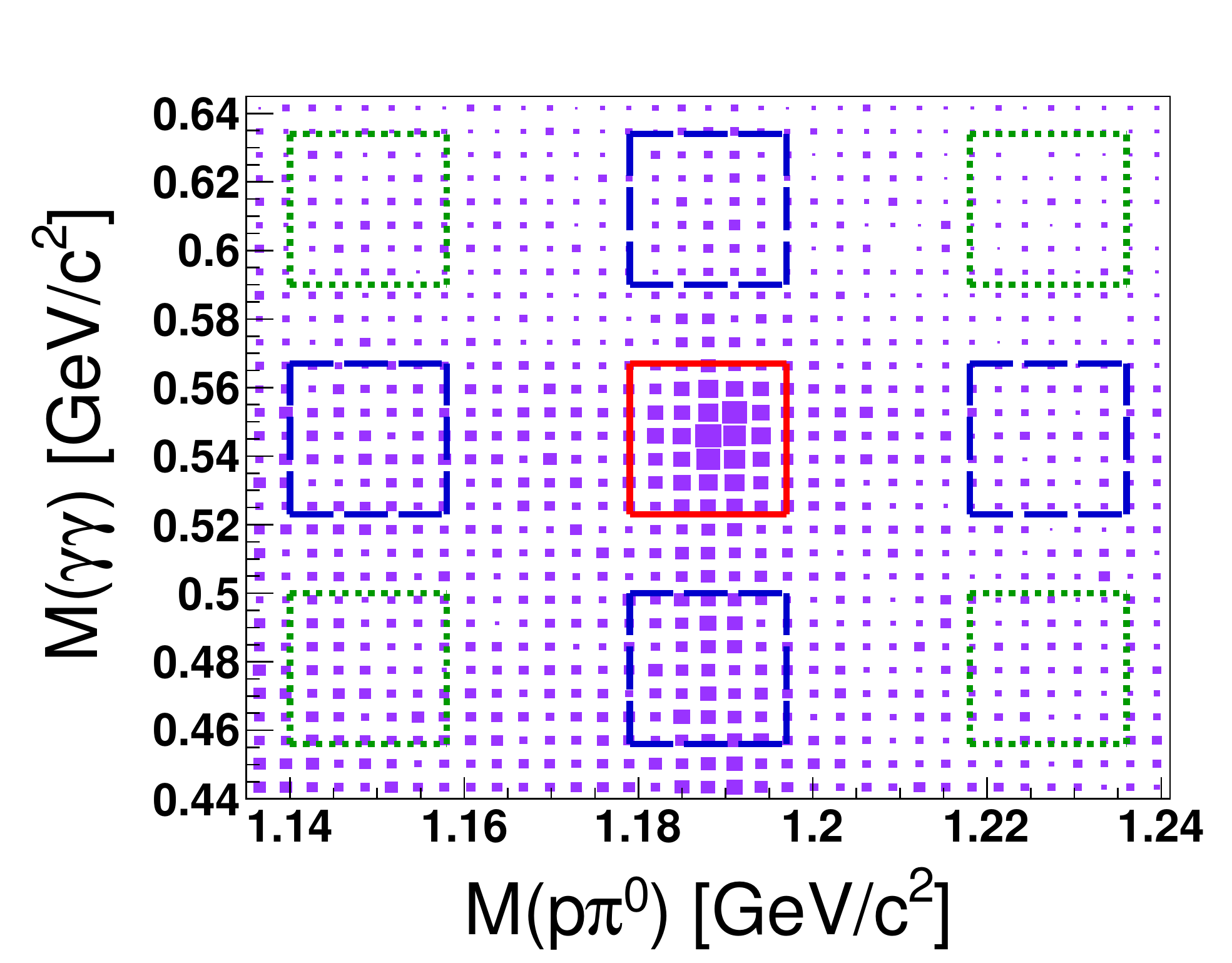}
              \includegraphics[width=5.9cm]{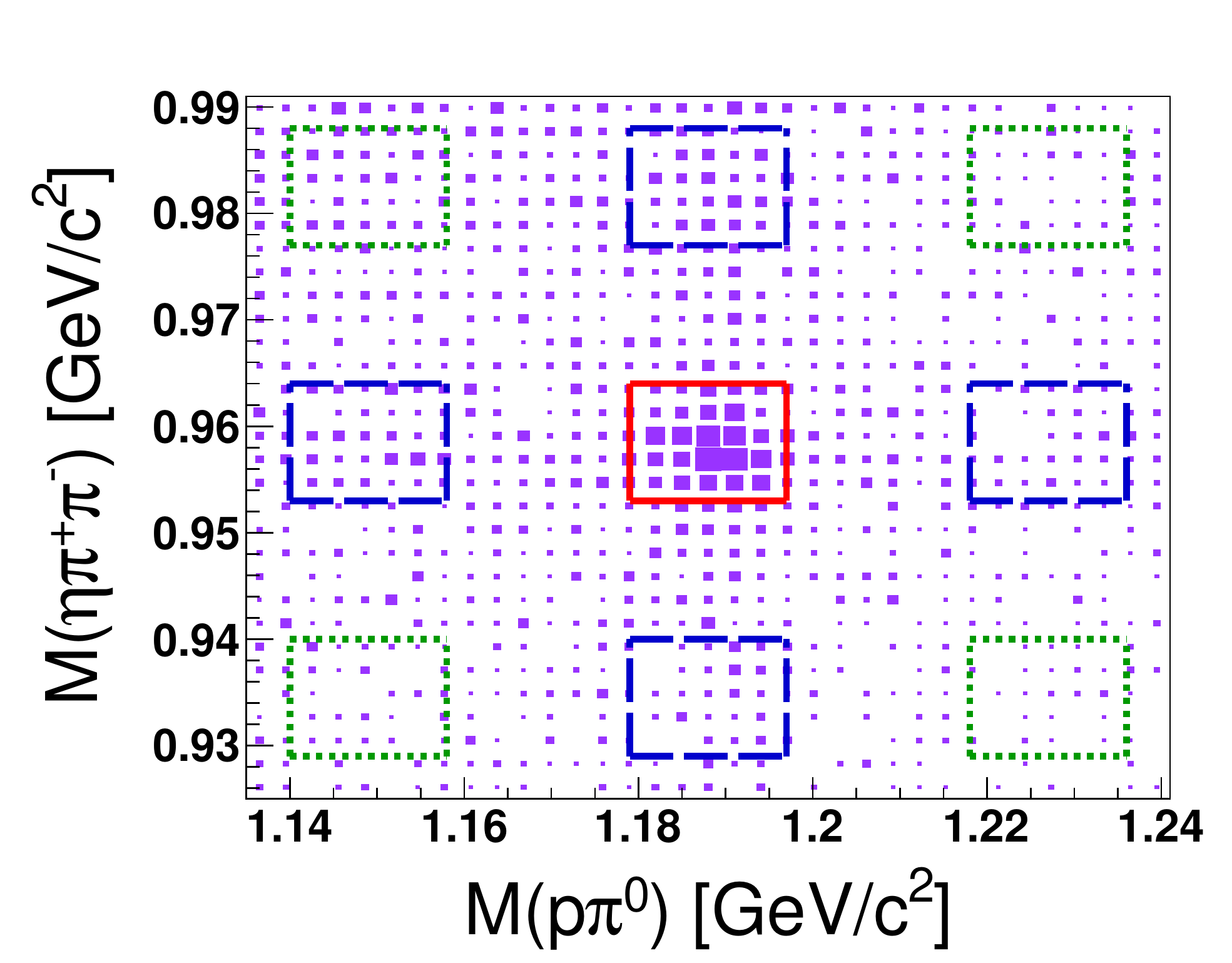}
              \put(-397,100){\large \bf (a)} \put(-227,100){\large \bf (b)}  \put(-55,100){\large \bf (c)}
              \caption{Two-dimensional distributions of $M(p\pi^0)$ versus $M(\gamma\gamma)$ for (a) $\Lambda_c^+ \to \Sigma^+ \pi^0$ and (b) $\Lambda_c^+ \to \Sigma^+ \eta$, and of $M(p\pi^0)$ versus $M(\eta\pi^+\pi^-)$ for (c) $\Lambda_c^+ \to \Sigma^+ \eta'$ from data. The red box is selected as the $\Lambda_c^+$ signal region and the blue and green boxes are selected as sidebands.}
              \label{fig:sidebands}
           \end{center}
     \end{figure*}

     The two-dimensional distributions of $M(p\pi^0)$ versus $M(\gamma\gamma)$ for $\Sigma^+ \pi^0$ and $\Sigma^+ \eta$ modes, and of $M(p\pi^0)$ versus $M(\eta\pi^+\pi^-)$ for $\Sigma^+ \eta'$ mode from data are shown in Fig.~\ref{fig:sidebands}. The red box is selected as $\Lambda_c^+$ signal region and the blue and green boxes are selected as sidebands. The number of the normalized sideband events is calculated to be 50\% of the number of events in blue boxes minus 25\% of the number of events in green boxes. The $\Lambda_c^+$ mass spectra from the normalized sideband events are shown as the yellow histograms in Fig.~\ref{fig:invmass}. The broken-signal also contributes in sidebands and no other peaking components are found.
     The other background contributions from the inclusive MC samples after subtracting the signal processes, broken-signal processes, and the sidebands are shown in Fig.~\ref{fig:invmass} with dark cyan histograms. The processes of $e^+e^- \to \Sigma^+h^0+anything$ could contribute a smooth distribution in the $M(\Sigma^+h^0)$ spectrum, especially for the $\Sigma^+\eta'$ mode.
     The sum of sidebands in data and other backgrounds in inclusive MC samples can describe the backgrounds in $\Lambda_c^+$ signal region well, indicating that there are no other peaking components. It is verified by the analysis of the inclusive MC samples with TopoAna~\cite{topology}.

\section{\boldmath Branching fraction}
\begin{figure}[h!tbp]
           \includegraphics[width=8cm]{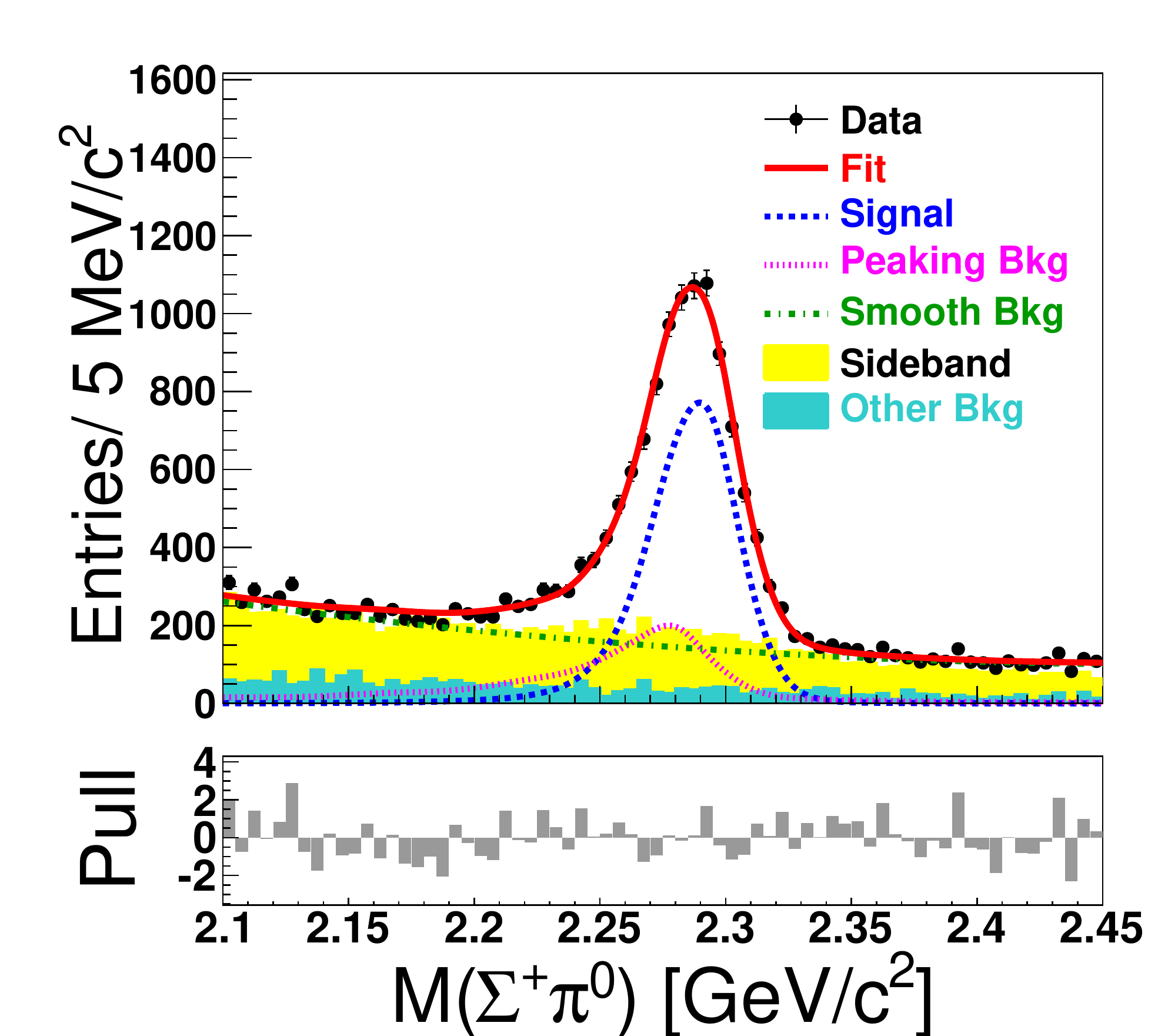}
           \put(-165, 150){\large \bf (a)}
           \vspace{0.0cm}

           \includegraphics[width=8cm]{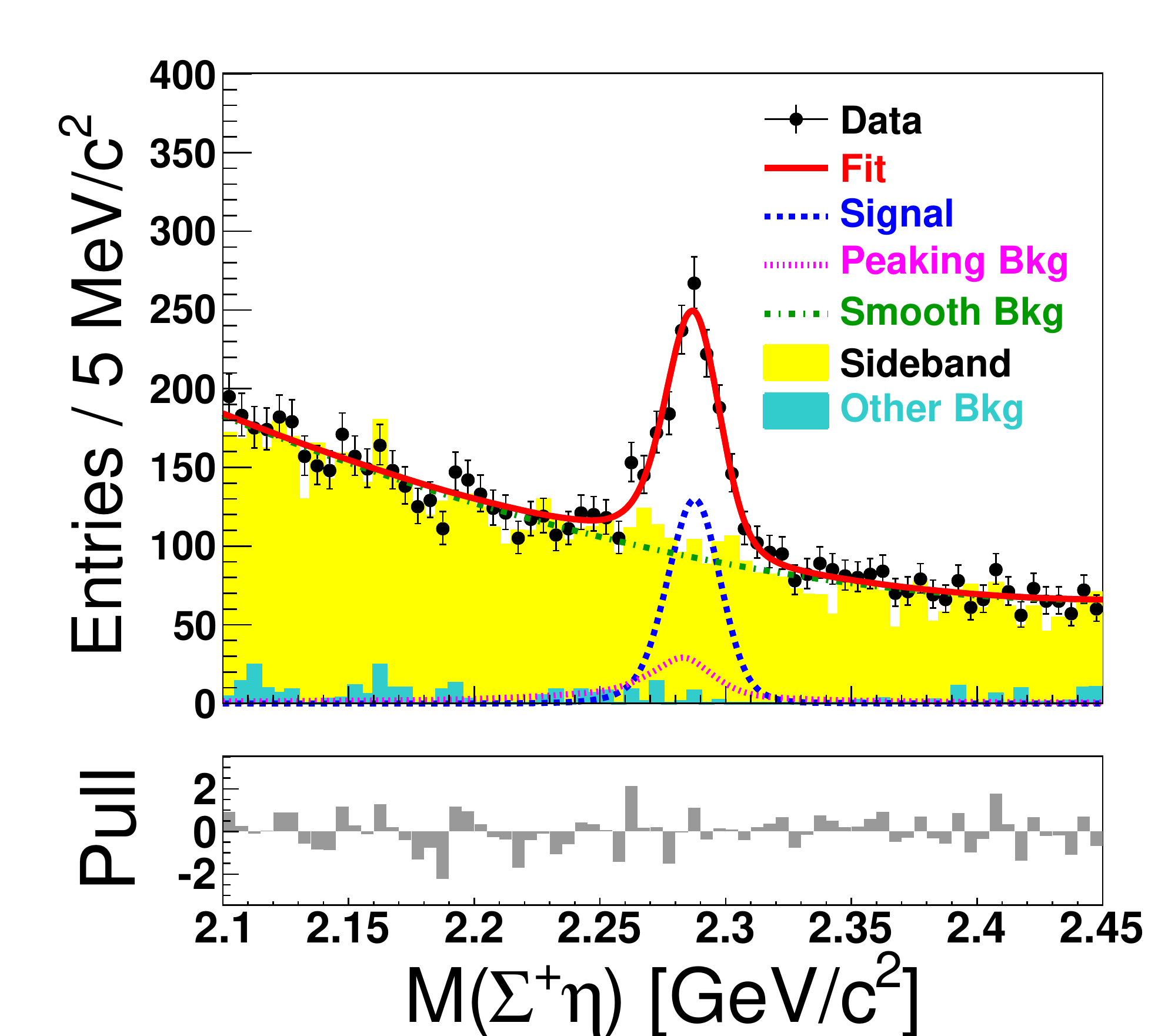}
           \put(-165, 150){\large \bf (b)}
           \vspace{0.0cm}

           \includegraphics[width=8cm]{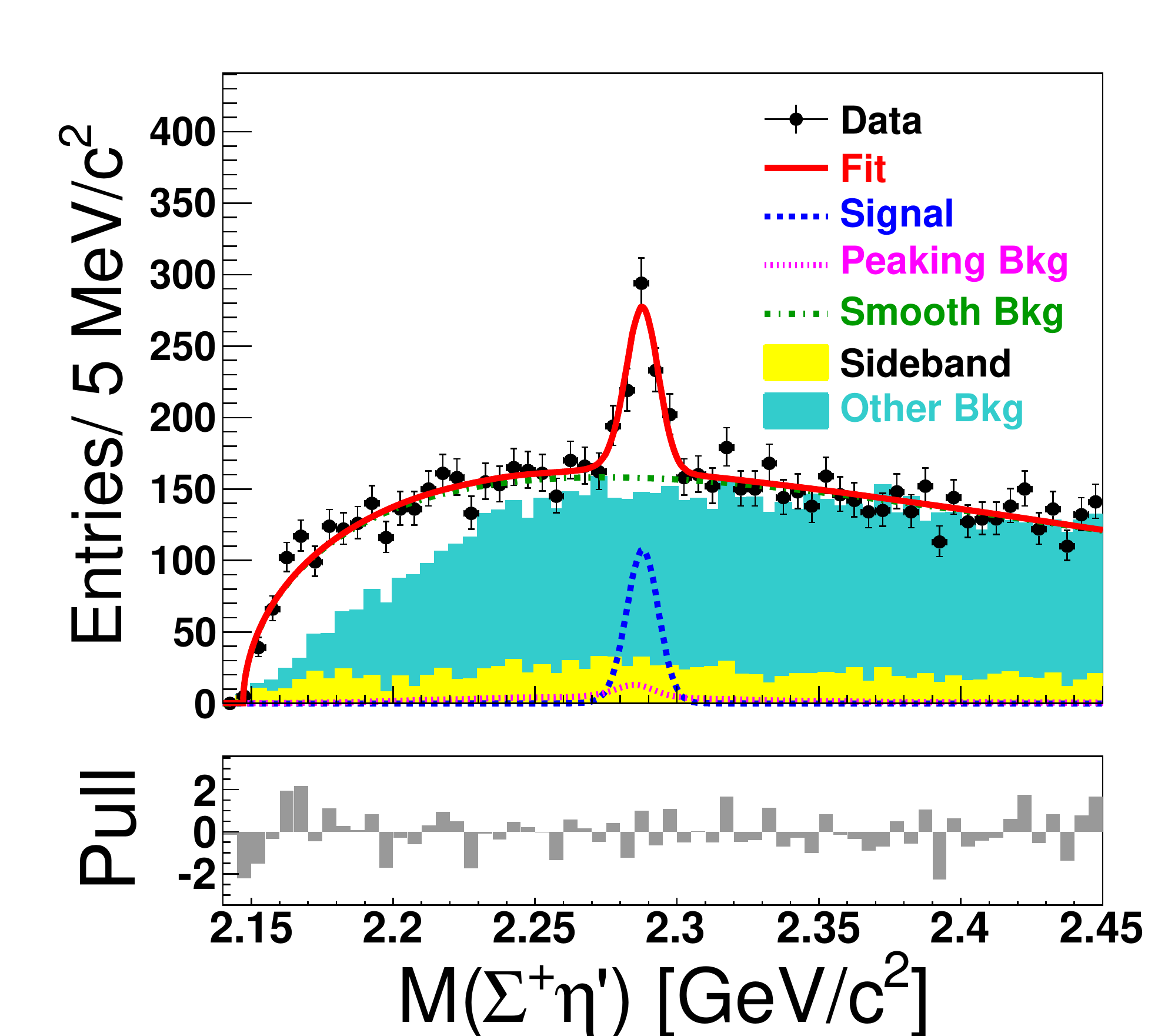}
           \put(-165, 150){\large \bf (c)}
           \caption{Invariant mass distributions of (a) $\Sigma^+ \pi^0$, (b) $\Sigma^+ \eta$, and (c) $\Sigma^+ \eta'$ from data. The black dots with error bars represent the data. The red solid line, blue dashed line, magenta dotted line, and dark green dash-dotted line stand for the best fit, signal shape, broken-signal shape, and the background shape, respectively. The yellow histogram is from the normalized sidebands and the dark cyan histogram is from the normalized inclusive MC samples after subtracting the signal process, broken-signal process, and sidebands.}
           \label{fig:invmass}
     \end{figure}

     After applying all the event selections, the invariant mass distributions of $\Sigma^+ \pi^0$, $\Sigma^+ \eta$, and $\Sigma^+ \eta'$ from data are shown in Fig.~\ref{fig:invmass}. The $\Lambda_c^+$ signal yields are extracted by performing an unbinned maximum-likelihood fit to these invariant-mass spectra.
     The likelihood function is defined in terms of the signal probability density function (PDF) $F_{\rm S}$, broken-signal PDF $F_{\rm B1}$,  and smooth background PDF $F_{\rm B2}$ as
     \begin{equation}
           \small
           \label{eq:likelihood}
           \mathcal{L}=\frac{e^{-(n_{\rm S}+n_{\rm B1}+n_{\rm B2})}}{N!}\prod_i^{N} \left[n_{\rm S}F_{\rm S}+n_{\rm B1}F_{\rm B1}+n_{\rm B2}F_{\rm B2}\right],
     \end{equation}
     where $N$ is the total number of observed events; $n_{\rm S}$, $n_{\rm B1}$, and $n_{\rm B2}$ are the numbers of $\Lambda_c^+$ signal events, broken-signal events, and smooth background events, respectively; $i$ denotes the event index. The numbers of $n_{\rm S}$ and $n_{\rm B2}$ are floated in the fit. Since the shapes of broken-signal events from the signal region and sidebands agree, we let the number of broken-signal events free in the fit.
     The $F_{\rm S}$ PDF is modeled with a histogram from the MC simulated events containing the true $\Lambda_c^+$ signal using rookeyspdf~\cite{keyspdf} and convolved with a Gaussian function, of which the mean and resolution are floated. The Gaussian describes the detector resolution difference between data and MC simulation. The $F_{\rm B1}$ PDF is smoothed with rookeyspdf from the MC simulation to describe the shape of broken-signal contribution. The $F_{\rm B2}$ PDF is parameterized by a first-order polynomial for the $\Sigma^+ \pi^0$ mode, a second-order polynomial for the $\Sigma^+ \eta$ mode, and a threshold function for the $\Sigma^+ \eta'$ mode. We define the threshold function as $(M-m_0)^\alpha  \rm exp(-\it \beta(M-m_{\rm 0}))$, where $M$ is the invariant mass of $\Sigma^+ \eta'$ and the parameter $m_0$ is the sum of the nominal masses of $\Sigma^+$ and $\eta'$~\cite{pdg}. We let the parameters $\alpha$ and $\beta$ float and $m_0$ be fixed in the fit. The best fit results for the three modes are shown in Fig.~\ref{fig:invmass}, along with the pulls $(N_{\rm data} - N_{\rm fit})/\sigma_{\rm data}$, where $\sigma_{\rm data}$ is the error on $N_{\rm data}$.

     An ensemble test comprising 1000 pseudoexperiments, where the PDF shapes from fit to data are used, is performed to check for any bias in the fit. We obtain a Gaussian normalized pull distribution width and mean consistent with one and zero within statistical uncertainties, respectively, indicating no fit bias.

     The $\Lambda_c^+$ signal yields and the signal efficiencies estimated from the signal MC simulations for three modes are summarized in Table~\ref{tab:bf:Nfit:eff}.

     \begin{table}[h]
             \centering
             \caption{The $\Lambda_c^+$ signal yields and the signal efficiencies for different decay modes.}
             \label{tab:bf:Nfit:eff}
             \setlength{\tabcolsep}{2.mm}{
             \begin{tabular}{ c  c  c  c }
                        \hline \hline
                         Mode  & $\lamc \to \sgmpiz$  & $\lamc \to \sgmeta$ & $\lamc \to \sgmetp$ \\ \hline \specialrule{0em}{1pt}{1pt}
                         Signal yield    &$6965\pm142$  &$719\pm78$    &$307\pm55$    \\
                         Efficiency (\%)  &$0.63$        &$0.67$        &$0.50$ \\ \hline
             \end{tabular}
             }
    \end{table}

     The ratios of branching fractions are obtained via
     \begin{align}
          \label{equ-branching}
          \small
          \frac{\BF_{\lamc \to \sgmeta}}{\BF_{\lamc \to \sgmpiz}} &= \frac{(N_{\sgmeta}/\epsilon_{\sgmeta})}{(N_{\sgmpiz}/\epsilon_{\sgmpiz})}\frac{\BF_{\pi^0\to\gamma\gamma}}{\BF_{\eta\to\gamma\gamma}},\\
          \frac{\BF_{\lamc \to \sgmetp}}{\BF_{\lamc \to \sgmpiz}} &= \frac{(N_{\sgmetp}/\epsilon_{\sgmetp})}{(N_{\sgmpiz}/\epsilon_{\sgmpiz})}\frac{\BF_{\pi^0\to\gamma\gamma}}{\BF_{\eta' \to \eta\pi^+\pi^-}\BF_{\eta \to \gamma\gamma}},\\
          \frac{\BF_{\lamc \to \sgmetp}}{\BF_{\lamc \to \sgmeta}} &= \frac{(N_{\sgmetp}/\epsilon_{\sgmetp})}{(N_{\sgmeta}/\epsilon_{\sgmeta})}\frac{1}{\BF_{\eta' \to \eta\pi^+\pi^-}},
     \end{align}
     where $N_{\sgmeta}$, $N_{\sgmetp}$, and $N_{\sgmpiz}$ are the $\Lambda_c^+$ signal yields for the three modes; $\epsilon_{\sgmeta}$, $\epsilon_{\sgmetp}$, and $\epsilon_{\sgmpiz}$ are the signal efficiencies; $\BF_{\pi^0/\eta\to\gamma\gamma}$ and $\BF_{\eta' \to \eta\pi^+\pi^-}$ are the branching fractions of decays of the intermediate states. We obtain
     \begin{align}
     \small
     \frac{\BF_{\lamc \to \sgmeta}}{\BF_{\lamc \to \sgmpiz}} &= 0.25 \pm 0.03 \pm 0.01,\\
     \frac{\BF_{\lamc \to \sgmetp}}{\BF_{\lamc \to \sgmpiz}} &= 0.33 \pm 0.06 \pm 0.02, \\
     \frac{\BF_{\lamc \to \sgmetp}}{\BF_{\lamc \to \sgmeta}} &= 1.34 \pm 0.28 \pm 0.06,
     \end{align}
     where the first uncertainty is statistical and the second systematic, as discussed in Section~\ref{sec:systematic:bf}.

\section{\boldmath asymmetry parameter}
     The differential decay rate depends on the asymmetry parameter $\alpha_{\Sigma^+ h^0}$ as
     \begin{equation}
     \label{equ:alpha}
      \frac{dN}{d\cos\theta_{\Sigma^+}} \propto 1+ \alpha_{\Sigma^+ h^0}\alpha_{p\pi^0}\cos\theta_{\Sigma^+}.
     \end{equation}

     We divide the $\cos\theta_{\Sigma^+}$ distribution into five bins. The $\Lambda_c^+$ signal yield in each bin is obtained by fitting the corresponding $M(\Sigma^+ h^0)$ distribution with a method similar to that used in the overall fit. The fits to the $M(\Sigma^+ h^0)$ distributions in each bin of $\cos\theta_{\Sigma^+}$ are shown in Appendix. Due to limited statistics, the fraction of $n_{\rm S}$ to $n_{\rm B1}$ and the values of the mean and resolution of the Gaussian function are fixed to those from overall fits for $\Sigma^+ \eta$ and $\Sigma^+ \eta'$ modes.
     No dependence of $n_{\rm S}/n_{\rm B1}$ on $\cos\theta_{\Sigma^+}$ is found in data for the $\Sigma^+\pi^0$ mode, and none is found in MC simulation for the $\Sigma^+ \eta$ and $\Sigma^+ \eta'$ modes, so fixing this ratio from MC simulation is reasonable.
     Table~\ref{tab:alpha:massfit} lists the signal yields and signal efficiencies in the five $\cos\theta_{\Sigma^+}$ bins for the three decays.

     \begin{table}[h]
             \centering
             \footnotesize
             \caption{The values of $\frac{\rm Signal~yields}{\rm Efficiency (\%)}$ in different $\cos\theta_{\Sigma^+}$ bins for $\sgmpiz$, $\sgmeta$, and $\sgmetp$ modes.}
             \begin{tabular}{c c c c  c c  } \hline \hline
               $\cos\theta_{\Sigma^+}$ & ($-1$, $-0.6$) & ($-0.6$, $-0.2$) & ($-0.2$, 0.2) & (0.2, 0.6) & (0.6, 1) \\ \hline
               \specialrule{0em}{2pt}{2pt}
               $\sgmpiz$ &$\frac{1019\pm98}{0.69}$ & $\frac{1240\pm104}{0.67}$ & $\frac{1350\pm117}{0.62}$ & $\frac{1619\pm129}{0.59}$ & $\frac{1967\pm131}{0.61}$ \\
               \specialrule{0em}{2pt}{2pt}
               $\sgmeta$ &$\frac{44\pm20}{0.75}$ & $\frac{69\pm21}{0.72}$ & $\frac{129\pm27}{0.67}$ & $\frac{223\pm32}{0.67}$ & $\frac{256\pm35}{0.63}$ \\
               \specialrule{0em}{2pt}{2pt}
               $\sgmetp$ &$\frac{46\pm18}{0.51}$ & $\frac{47\pm17}{0.51}$ & $\frac{62\pm19}{0.49}$ & $\frac{54\pm18}{0.50}$ & $\frac{102\pm21}{0.49}$ \\
               \specialrule{0em}{1pt}{1pt} \hline \hline
              \end{tabular}
              \label{tab:alpha:massfit}
      \end{table}

\begin{figure}[h!tbp]
           \includegraphics[width=8cm]{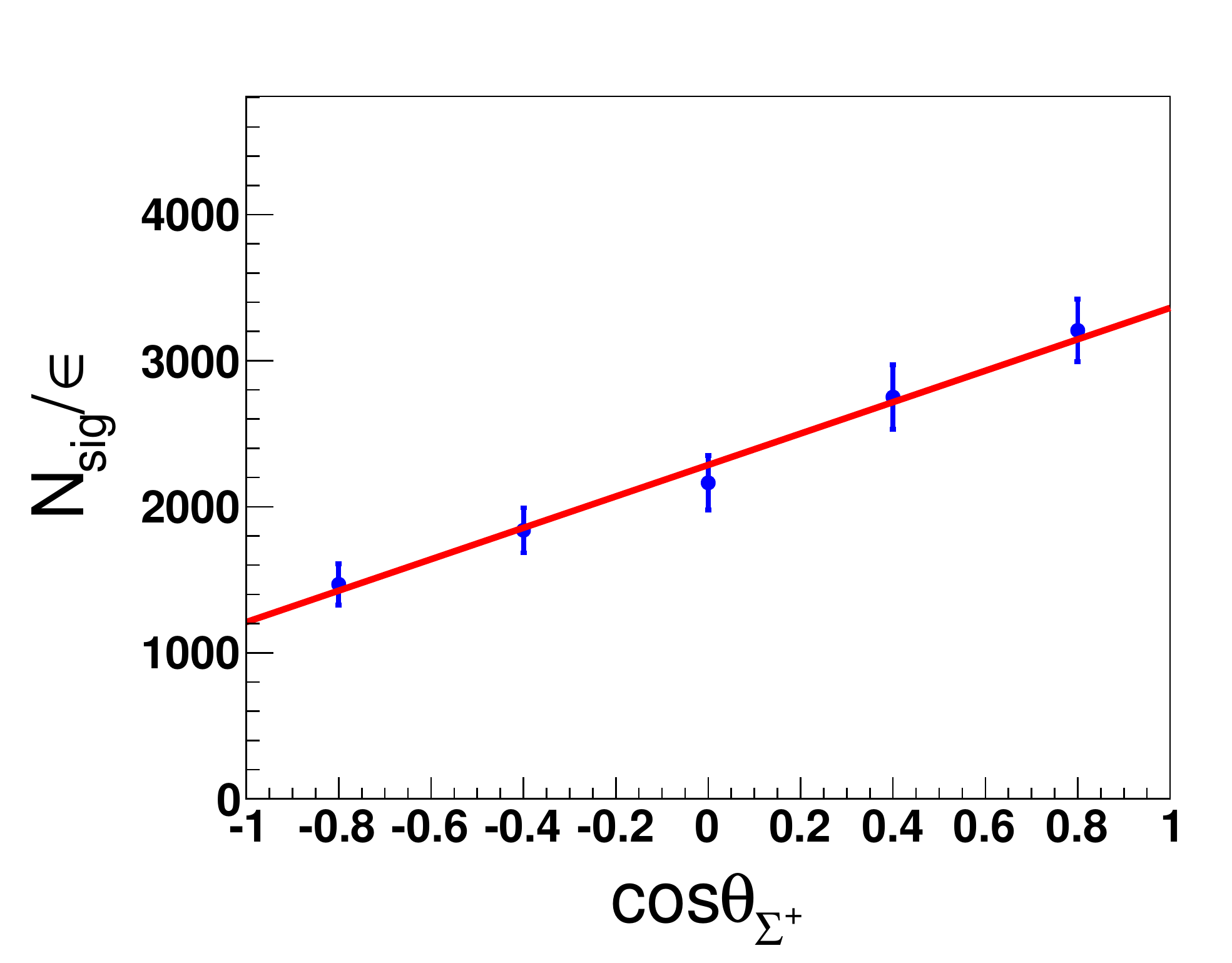}
           \put(-165, 140){\large \bf (a) $\lamc \to \sgmpiz$}
           \vspace{0.0cm}

           \includegraphics[width=8cm]{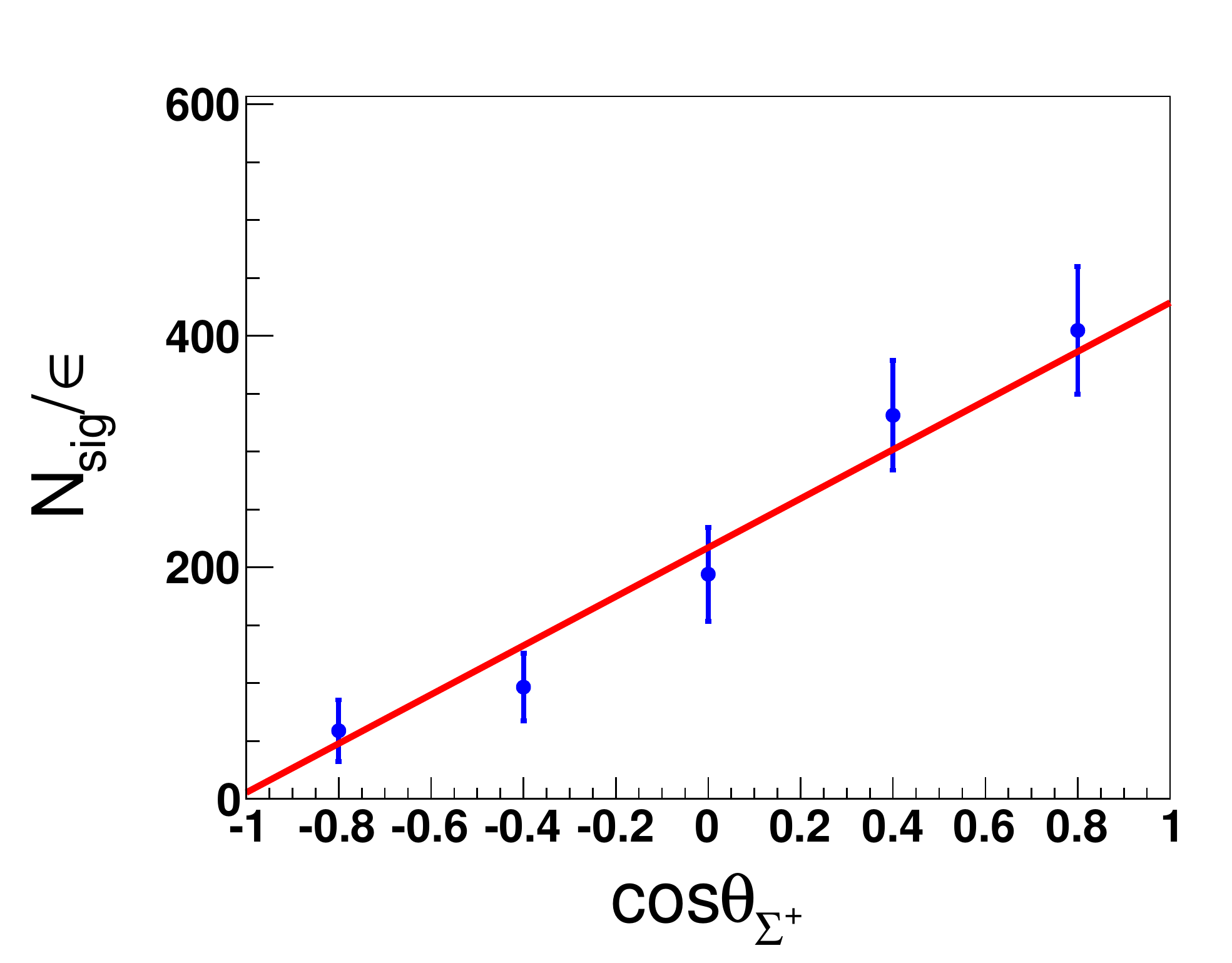}
           \put(-165, 140){\large \bf (b) $\lamc \to \sgmeta$}
           \vspace{0.0cm}

           \includegraphics[width=8cm]{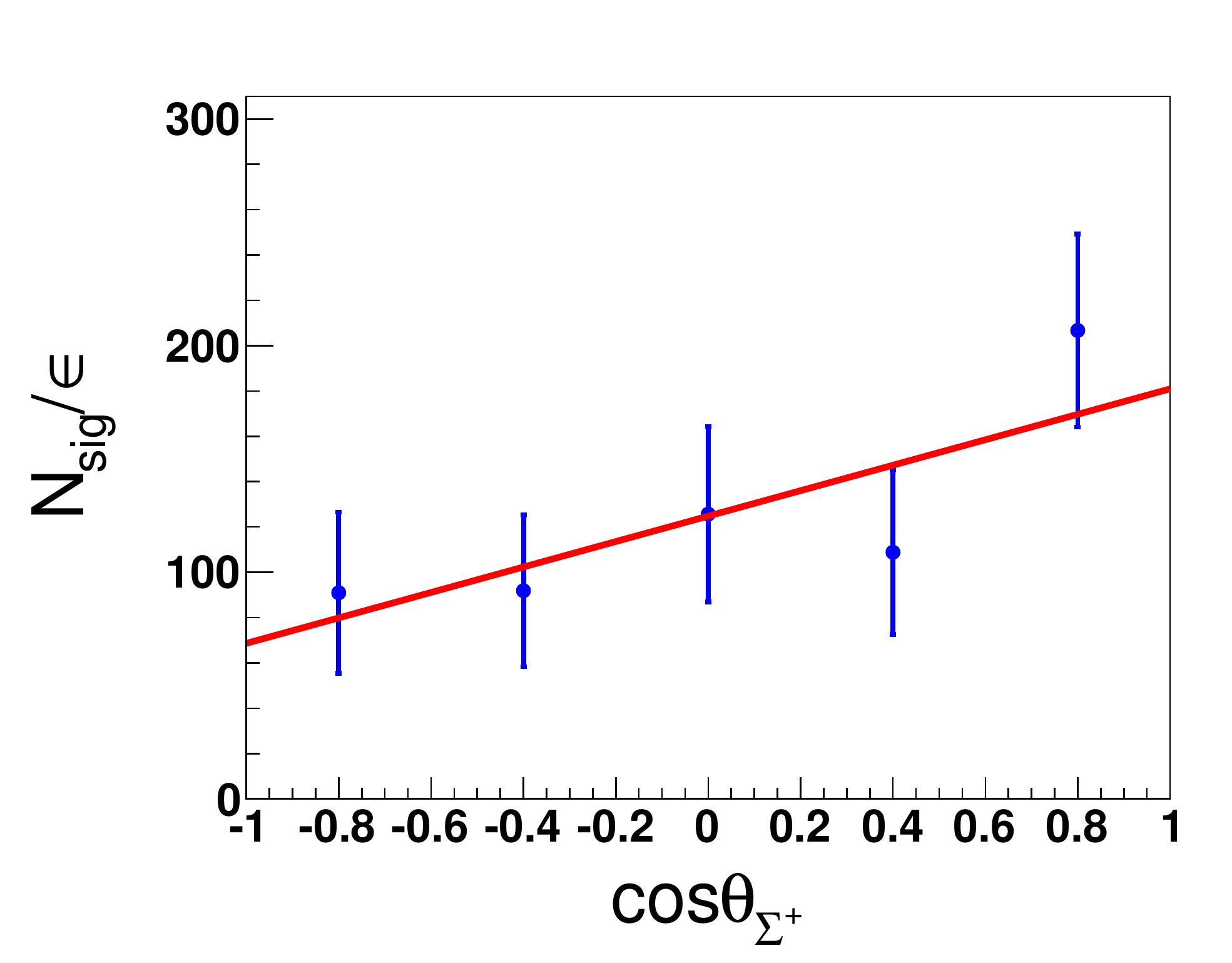}
           \put(-165, 140){\large \bf (c) $\lamc \to \sgmetp$}
           \caption{The maximum likelihood fits to the efficiency-corrected $\cos\theta_{\Sigma^+}$ distributions of data to extract (a) $\alpha_{\sgmpiz}$, (b) $\alpha_{\sgmeta}$, and (c) $\alpha_{\sgmetp}$. The points with error bars represent data and the red solid lines are the best fits.}
           \label{fig:alpha:linear}
     \end{figure}

      The final efficiency-corrected $\cos\theta_{\Sigma^+}$ distributions for $\lamc \to \sgmpiz$, $\lamc \to \sgmeta$, and $\lamc \to \sgmetp$ decays are shown in Fig.~\ref{fig:alpha:linear}. Using the function $dN/d\cos\theta_{\Sigma^+}\propto 1+ \alpha_{\Sigma^+ h^0}\alpha_{p\pi^0}\cos\theta_{\Sigma^+}$ to fit the efficiency-corrected $\cos\theta_{\Sigma^+}$ distributions, where the parameter $\alpha_{\Sigma^+ h^0}\alpha_{p\pi^0}$ is floating in the fit, we obtain
      \begin{align}
                \alpha_{\sgmpiz}\alpha_{p\pi^0} &= 0.47 \pm 0.02,\\
                \alpha_{\sgmeta}\alpha_{p\pi^0} &= 0.97 \pm 0.03 ,\\
                \alpha_{\sgmetp}\alpha_{p\pi^0} &= 0.45 \pm 0.06 ,
           \end{align}
     where the uncertainties are statistical. Using $\alpha_{p\pi^0}=-0.983 \pm 0.013$~\cite{pdg}, we calculate the asymmetry parameters
     \begin{align}
          \alpha_{\sgmpiz} &= -0.48 \pm 0.02 \pm 0.02,\\
          \alpha_{\sgmeta} &= -0.99 \pm 0.03 \pm 0.05,\\
          \alpha_{\sgmetp} &= -0.46 \pm 0.06 \pm 0.03,
     \end{align}
     where the uncertainties are statistical and systematic, respectively. Section~\ref{sec:systematic:alpha} discusses the details of systematic uncertainty.

\section{\boldmath Systematic Uncertainties}
\subsection{\boldmath Ratio of the branching fractions}
\label{sec:systematic:bf}
     Systematic uncertainties in the ratios of branching fractions are summarized in Table~\ref{tab:syserr:bf}.
     The sources include particle identification (PID), charged-track reconstruction, mass windows of intermediate states, fit range, background PDF, signal PDF, broken-signal PDF, the number of broken-signal events, asymmetry parameter, branching fractions of intermediate states, and MC statistics. The systematic uncertainty from $\Sigma^+ \to p\pi^0$ reconstruction cancels for the three ratios, including the efficiency of proton selection and reconstruction, photon reconstruction, and mass window of $\Sigma^+$.

     Based on the study of $D^{*+} \to D^0( \to K^-\pi^+)\pi^+$, the PID uncertainty is 1.2\% for $\pi^+$ selection, and 1.3\% for $\pi^-$ selection. We use partially reconstructed $D^{*+}$ decays in $D^{*+} \to \pi^+ D^0 (\to K_S^0 \pi^+ \pi^-)$ to assign the systematic uncertainty from charged-track reconstruction (0.35\% per track).

     The ratios of the efficiencies between data and MC simulations are $0.986\pm0.004$, $1.009\pm0.008$, and $0.989\pm0.011$ for mass windows of $\pi^0$, $\eta$, and $\eta'$, respectively. The central values of the ratios are taken as the efficiency correction factors and the uncertainties are taken as systematic uncertainties due to mass windows of the intermediate states.

     We perform the simulated pseudoexperiments to estimate the uncertainty from the fit procedure. An ensemble of 1000 pseudoexperiments is generated using the mass distributions of $\lamc$ in Fig.~\ref{fig:invmass} and fitted with a changed likelihood function. The distribution of the fitted $\lamc$ signal yields is fitted by a Gaussian function and the difference between the mean of Gaussian and the nominal value is taken as the systematic uncertainty.  The changes of fit condition include (1) changing the fit range to (2.15, 2.40) GeV/$c^2$; (2) changing the background PDF from a first-order polynomial to a second-order one for $\lamc \to \sgmpiz$, from a second-order polynomial to a third-order one for $\lamc \to \sgmeta$, and from a threshold function to an ARGUS function~\cite{argus} for $\lamc \to \sgmetp$; (3) using the signal shape smoothed by roohistpdf~\cite{root}; (4) using the broken-signal shape smoothed by roohistpdf. We add the uncertainties in quadrature for the corresponding ratios of branching fractions.

     The number of broken-signal events is floating in the nominal fit to account for the contribution from broken-signal in sidebands. Considering the possibility of over-estimation or under-estimation of the broken-signal contribution, we fix the number of broken-signal events according to the MC simulation, and take the difference in ratio of branching fractions as the systematic uncertainty.

     The signal MC samples are corrected using the measured asymmetry parameters. We vary the asymmetry parameter by $\pm 1\sigma$ and take the maximum difference on the signal efficiency with respect to the nominal value as the systematic uncertainty, which is 0.4\% for the $\sgmpiz$ mode, 1.7\% for the $\sgmeta$ mode, and 0.3\% for the $\sgmetp$ mode. We add them in quadrature for the corresponding ratios of branching fractions.

     The systematic uncertainties from $\BF_{\pi^0 \to \gamma\gamma}$, $\BF_{\eta' \to \eta\pi^+\pi^-}$, and $\BF_{\eta \to \gamma\gamma}$ are 0.03\%, 1.2\%, and 0.5\%, respectively~\cite{pdg}. The systematic uncertainty due to MC statistics in the signal efficiency is calculated as $\sqrt{(1 - \epsilon)/(\epsilon N_{\rm gen}})$, where $\epsilon$ is the signal efficiency and $N_{\rm gen}$ is the number of generated $\lamc$ simulated events.

     \begin{table}[h]
             \centering
             \caption{Relative systematic uncertainties (in \%) in ratios of branching fractions.}
             \label{tab:syserr:bf}
             \begin{tabular}{ c  c  c  c }
                        \hline \hline
                         Source  & $\frac{\BF_{\lamc \to \sgmeta}}{\BF_{\lamc \to \sgmpiz}}$  & $\frac{\BF_{\lamc \to \sgmetp}}{\BF_{\lamc \to \sgmpiz}}$ & $\frac{\BF_{\lamc \to \sgmetp}}{\BF_{\lamc \to \sgmeta}}$ \\
                         \hline
                         \specialrule{0em}{2pt}{2pt}
                         PID      &-- &2.5 &2.5    \\
                         Track reconstruction &-- &0.7 &0.7 \\
                         Mass window &0.9 &1.4 &1.1 \\
                         Fit range      &1.9 &1.2 &1.6 \\
                         Background PDF &1.4 &3.0 &2.9 \\
                         Signal PDF        &0.5 &0.2 &0.7 \\
                         Broken-signal PDF &1.2 &1.1 &0.8 \\
                         Number of broken-signal &4.2 &3.6 &2.5 \\
                         Asymmetry parameter &1.7 &0.5 &1.7 \\
                         Branching fraction &0.5 &1.3 &1.2 \\
                         MC statistics &1.4 &1.5 &1.5 \\ \hline
                         Total    &5.6 &6.1 &5.7 \\ \hline \hline
             \end{tabular}
     \end{table}

\subsection{\boldmath asymmetry parameter}
\label{sec:systematic:alpha}
     Systematic uncertainties in asymmetry parameters are summarized in Table~\ref{tab:syserr:alpha}.
     The sources include mass resolution, fit range, background PDF, signal PDF, broken-signal PDF, the number of broken-signal events, the number of $\cos\theta_{\Sigma^+}$ bins, and $\alpha_{p\pi^0}$.

     The systematic uncertainties from the fit are estimated using simulated pseudoexperiments. We use an ensemble of pseudoexperiments to generate the mass spectra of $\lamc$ candidates from data samples. The number of signal events in each $\cos\theta_{\Sigma^+}$ bin is obtained by fitting the generated mass spectra after enlarging the mass resolution by 10\%, and changing the fit range, background shape, signal shape, and broken-signal shape. After 1000 simulations, distributions of $\alpha_{\Sigma^+ h^0}$ are obtained by fitting the slopes of the $\cos\theta_{\Sigma^{+}}$ distributions. The differences between the fitted values of the distributions of $\alpha_{\Sigma^+ h^0}$ with Gaussian functions and the nominal values are taken as the uncertainties. The changed fit range, background shape, signal shape, and broken-signal shape follow those in Section~\ref{sec:systematic:bf}. The parameters of the Gaussian describing the resolution difference between data and MC simulation are not fixed in $\alpha_{\sgmpiz}$ extraction, thus no uncertainty from mass resolution is considered.

     The number of broken-signal events is floating for $\Sigma^+\pi^0$ mode and fixed for another two modes. To estimate the uncertainty from the number of broken-signal events, we fix the number based on the MC simulation, and take the difference in asymmetry parameters as the systematic uncertainty.

     We change the number of $\cos\theta_{\Sigma^{+}}$ bins from five to four or to six, and take the maximum difference on the asymmetry parameter as the systematic uncertainty. The systematic uncertainty from $\alpha_{p\pi^0}$ is 1.3\%~\cite{pdg}.

     \begin{table}[h]
             \centering
             \caption{Relative systematic uncertainties (in \%) in asymmetry parameters.}
             \label{tab:syserr:alpha}
             \setlength{\tabcolsep}{4.0mm}{
             \begin{tabular}{ c  c  c  c }
                        \hline \hline
                         Source  & $\alpha_{\sgmpiz}$ & $\alpha_{\sgmeta}$  & $\alpha_{\sgmetp}$ \\ \hline
                         Mass resolution  & -- &1.4 &2.9    \\
                         Fit range        &1.6 &2.7 &2.1     \\
                         Background shape &1.1 &1.9 &1.0 \\
                         Signal PDF        &0.3 &0.4 &0.1 \\
                         Broken-signal PDF &0.7 &1.0 &1.2 \\
                         Number of broken-signal &2.4 &-- & -- \\
                         $\cos\theta_{\Sigma^{+}}$ bins &1.9 &2.8 &4.7 \\
                         $\alpha_{p\pi^0}$ &1.3 &1.3 &1.3 \\ \hline
                         Total        &3.9 &4.8 &6.2 \\ \hline \hline
             \end{tabular}
             }
     \end{table}

\section{\boldmath summary}
     We analyze the $\Sigma^+ h^0$ final states to study $\lamc$ decays using the full Belle data corresponding to an integrated luminosity of 980 $\rm fb^{-1}$. The branching fractions of $\lamc \to \sgmeta$ and $\lamc \to \sgmetp$ relative to that of the $\lamc \to \sgmpiz$ decay mode are measured to be
     \begin{align}
     \small
     \frac{\BF_{\lamc \to \sgmeta}}{\BF_{\lamc \to \sgmpiz}} &= 0.25 \pm 0.03 \pm 0.01, \\
     \frac{\BF_{\lamc \to \sgmetp}}{\BF_{\lamc \to \sgmpiz}} &= 0.33 \pm 0.06 \pm 0.02,
     \end{align}
     where the uncertainties, here and below, are statistical and systematic, respectively. Taking $\BF_{\lamc \to \sgmpiz}=(1.25 \pm 0.10)\%$~\cite{pdg}, the absolute branching fractions are
     \begin{align}
     \small
     \BF_{\lamc \to \sgmeta} &= (3.14 \pm 0.35 \pm 0.17 \pm 0.25)\times 10^{-3}, \\
     \BF_{\lamc \to \sgmetp} &= (4.16 \pm 0.75 \pm 0.25 \pm 0.33)\times10^{-3},
     \end{align}
     where the third uncertainty is from $\BF_{\lamc \to \sgmpiz}$. The measured results are the most precise results to date and agree with earlier results~\cite{pdg} within two standard deviations.

     The ratio of $\BF_{\lamc \to \sgmetp}$ to $\BF_{\lamc \to \sgmeta}$ is determined to be
     \begin{align}
     \frac{\BF_{\lamc \to \sgmetp}}{\BF_{\lamc \to \sgmeta}} =  1.34\pm 0.28 \pm 0.08,
     \end{align}
     in which the value of the ratio is consistent with one within 1.2 standard deviations.

     The asymmetry parameters $\alpha_{\sgmeta}$ and $\alpha_{\sgmetp}$ are measured for the first time. The values are
     \begin{align}
          \alpha_{\sgmeta} &= -0.99 \pm 0.03 \pm 0.05,\\
          \alpha_{\sgmetp} &= -0.46 \pm 0.06 \pm 0.03.
     \end{align}
     The asymmetry parameter $\alpha_{\sgmpiz}$ is measured as
     \begin{align}
          \alpha_{\sgmpiz} &= -0.48 \pm 0.02 \pm 0.02,
     \end{align}
     which agrees with the world average value~\cite{pdg} with the uncertainty considerably improved from 20\% to 6\%. Comparing the measured result with $\alpha_{\Sigma^0 \pi^+}=-0.463 \pm 0.016 \pm 0.008$ as measured by Belle~\cite{theo13}, their agreement within one standard deviation shows consistency with the prediction from the isospin symmetry~\cite{theo11}.

     Comparing with the theoretical predictions summarized in Table~\ref{tab:bfalpha:theory:experiment}, none of the theoretical predictions could predict the $\BF_{\lamc \to \sgmetp}$ well and only the result using the method based on the Heavy Quark Effective Theory (HQEF)~\cite{theo8} is basically consistent with the measured $\BF_{\lamc \to \sgmeta}$. We also note that none of the methods is able to account for all three measured decay asymmetries.
     The approaches based on the HQEF~\cite{theo8} and the constituent quark model~\cite{theo10} could successfully predict the $\alpha_{\sgmeta}$. The measured $\alpha_{\sgmpiz}$ and $\alpha_{\sgmetp}$ are consistent with the predictions based on $SU(3)_F$ flavor symmetry~\cite{theo11} and spectator quark model~\cite{theo1}, respectively. The other approaches failed in predicting the asymmetry parameters of the above three modes.

\section*{ACKNOWLEDGMENTS}
This work, based on data collected using the Belle detector, which was
operated until June 2010, was supported by
the Ministry of Education, Culture, Sports, Science, and
Technology (MEXT) of Japan, the Japan Society for the
Promotion of Science (JSPS), and the Tau-Lepton Physics
Research Center of Nagoya University;
the Australian Research Council including grants
DP180102629, 
DP170102389, 
DP170102204, 
DE220100462, 
DP150103061, 
FT130100303; 
Austrian Federal Ministry of Education, Science and Research (FWF) and
FWF Austrian Science Fund No.~P~31361-N36;
the National Natural Science Foundation of China under Contracts
No.~11675166,  
No.~11705209;  
No.~11975076;  
No.~12135005;  
No.~12175041;  
No.~12161141008; 
Key Research Program of Frontier Sciences, Chinese Academy of Sciences (CAS), Grant No.~QYZDJ-SSW-SLH011; 
Project ZR2022JQ02 supported by Shandong Provincial Natural Science Foundation;
the Ministry of Education, Youth and Sports of the Czech
Republic under Contract No.~LTT17020;
the Czech Science Foundation Grant No. 22-18469S;
Horizon 2020 ERC Advanced Grant No.~884719 and ERC Starting Grant No.~947006 ``InterLeptons'' (European Union);
the Carl Zeiss Foundation, the Deutsche Forschungsgemeinschaft, the
Excellence Cluster Universe, and the VolkswagenStiftung;
the Department of Atomic Energy (Project Identification No. RTI 4002) and the Department of Science and Technology of India;
the Istituto Nazionale di Fisica Nucleare of Italy;
National Research Foundation (NRF) of Korea Grant
Nos.~2016R1\-D1A1B\-02012900, 2018R1\-A2B\-3003643,
2018R1\-A6A1A\-06024970, RS\-2022\-00197659,
2019R1\-I1A3A\-01058933, 2021R1\-A6A1A\-03043957,
2021R1\-F1A\-1060423, 2021R1\-F1A\-1064008, 2022R1\-A2C\-1003993;
Radiation Science Research Institute, Foreign Large-size Research Facility Application Supporting project, the Global Science Experimental Data Hub Center of the Korea Institute of Science and Technology Information and KREONET/GLORIAD;
the Polish Ministry of Science and Higher Education and
the National Science Center;
the Ministry of Science and Higher Education of the Russian Federation, Agreement 14.W03.31.0026, 
and the HSE University Basic Research Program, Moscow; 
University of Tabuk research grants
S-1440-0321, S-0256-1438, and S-0280-1439 (Saudi Arabia);
the Slovenian Research Agency Grant Nos. J1-9124 and P1-0135;
Ikerbasque, Basque Foundation for Science, Spain;
the Swiss National Science Foundation;
the Ministry of Education and the Ministry of Science and Technology of Taiwan;
and the United States Department of Energy and the National Science Foundation.
These acknowledgements are not to be interpreted as an endorsement of any
statement made by any of our institutes, funding agencies, governments, or
their representatives.
We thank the KEKB group for the excellent operation of the
accelerator; the KEK cryogenics group for the efficient
operation of the solenoid; and the KEK computer group and the Pacific Northwest National
Laboratory (PNNL) Environmental Molecular Sciences Laboratory (EMSL)
computing group for strong computing support; and the National
Institute of Informatics, and Science Information NETwork 6 (SINET6) for
valuable network support.

\section*{APPENDIX}

\begin{figure*}[htbp]
      \flushleft
          \includegraphics[width=5.5cm]{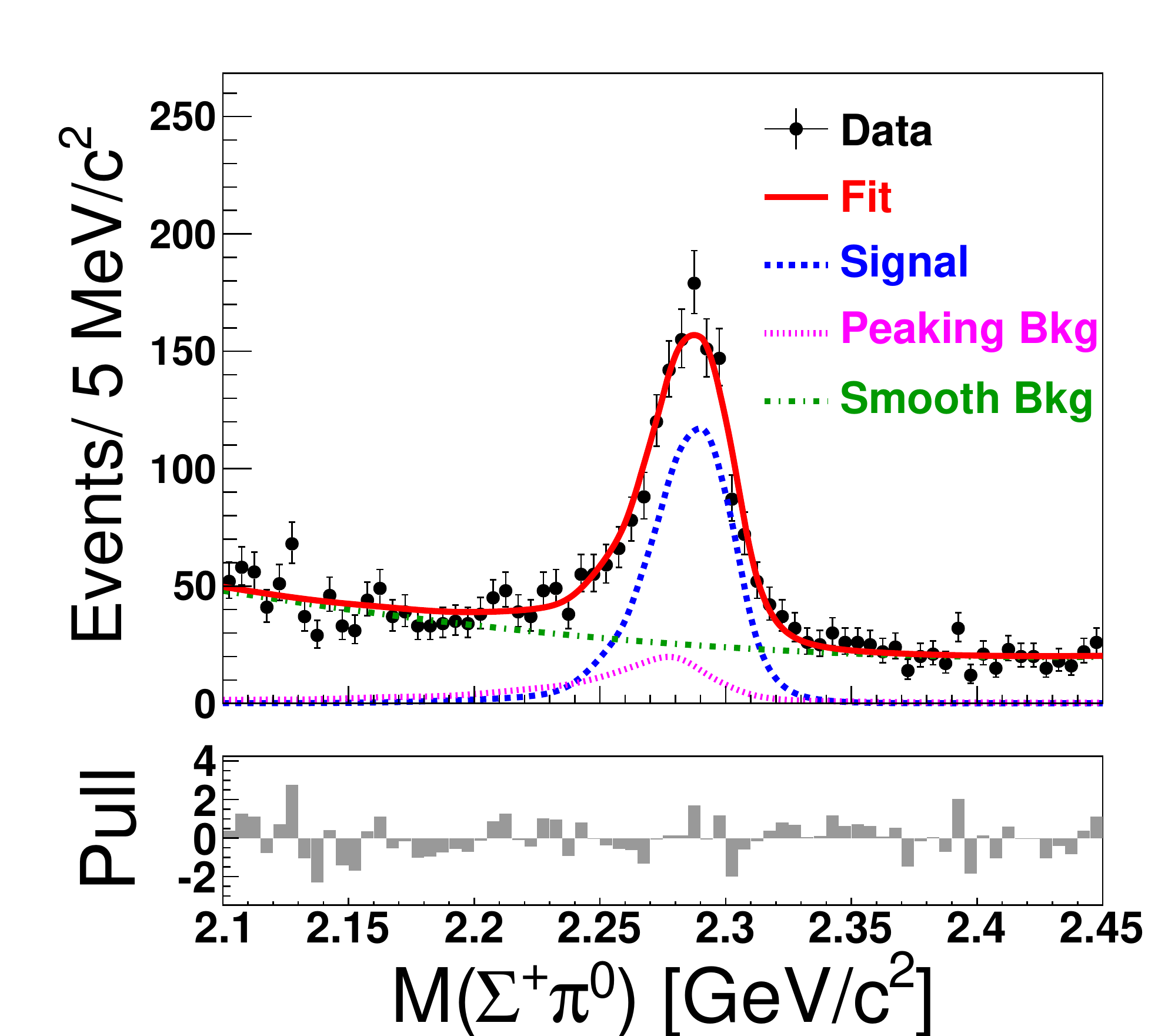}
          \includegraphics[width=5.5cm]{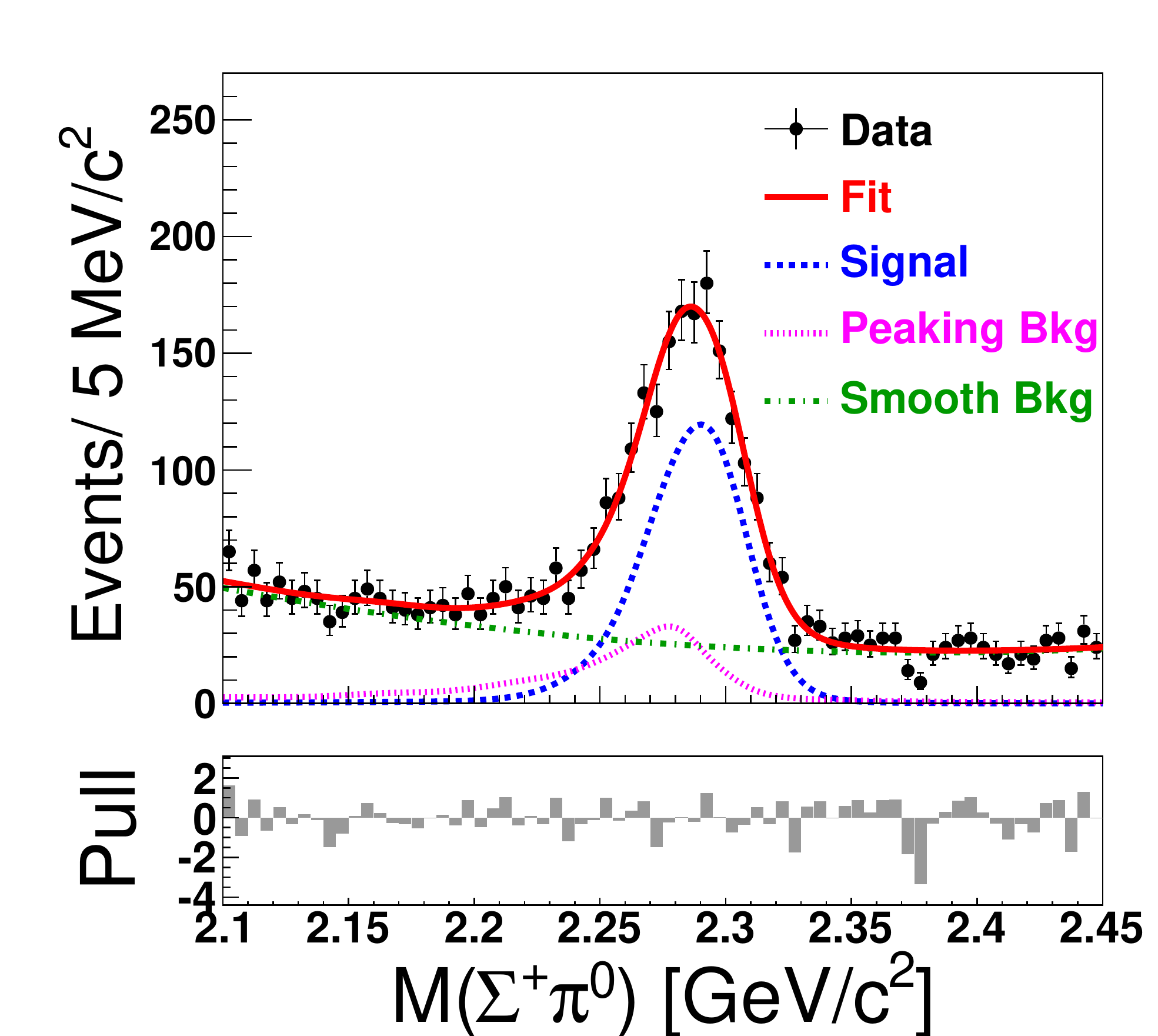}
          \includegraphics[width=5.5cm]{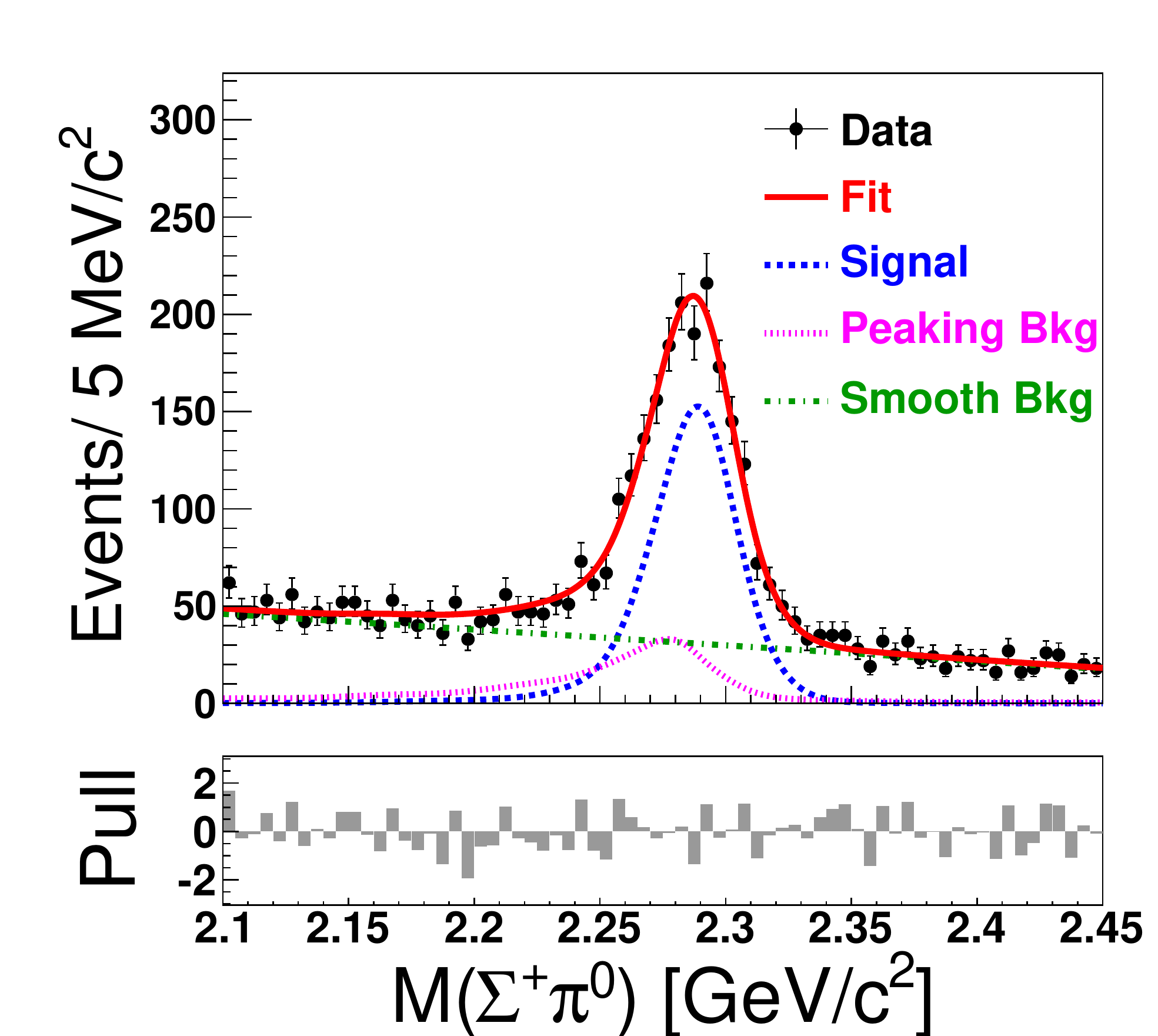}
          \put(-438,115){\large \bf (a)} \put(-277,115){\large \bf (b)}  \put(-115,115){\large \bf (c)}
          \quad
          \includegraphics[width=5.5cm]{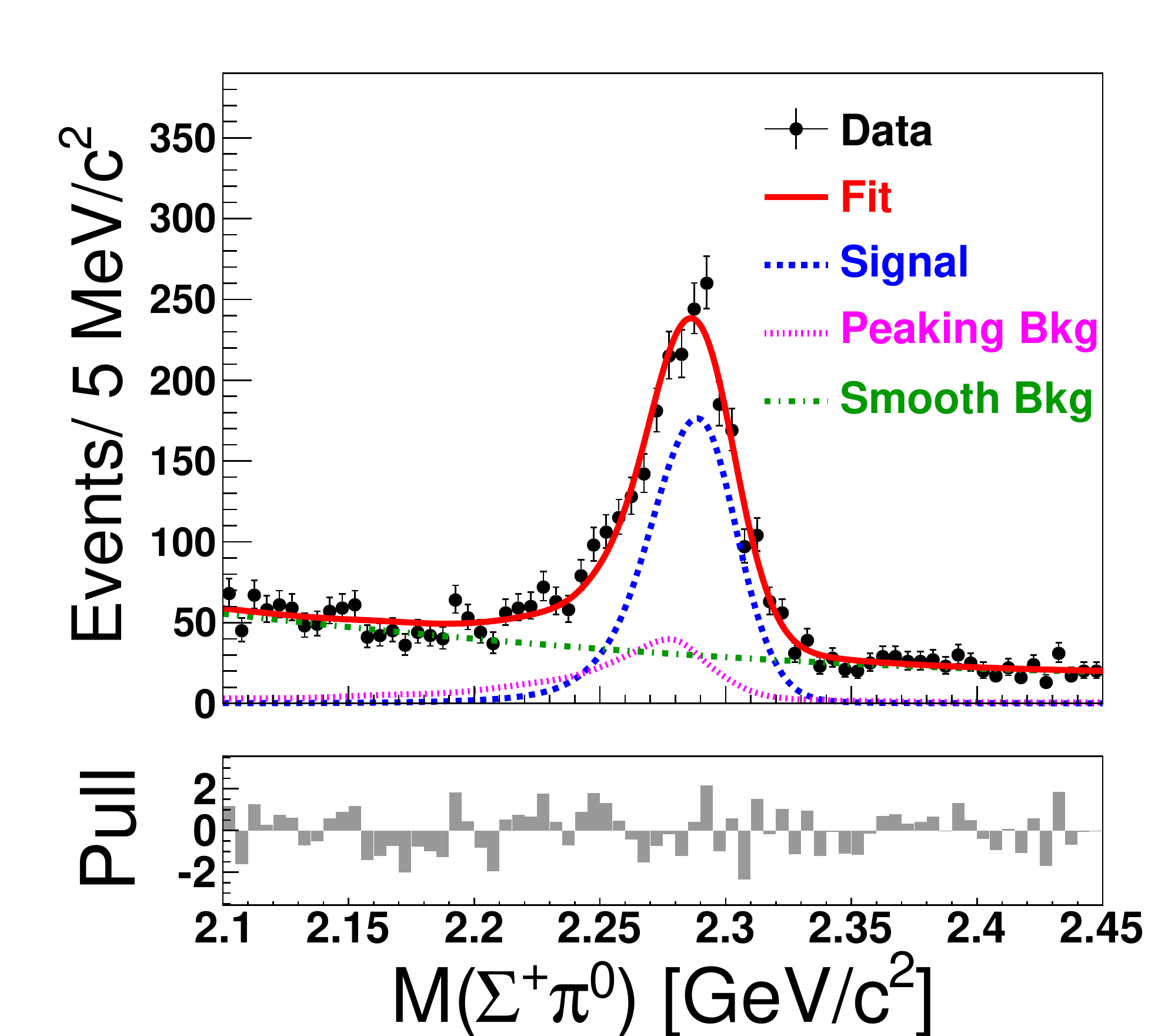}
          \includegraphics[width=5.5cm]{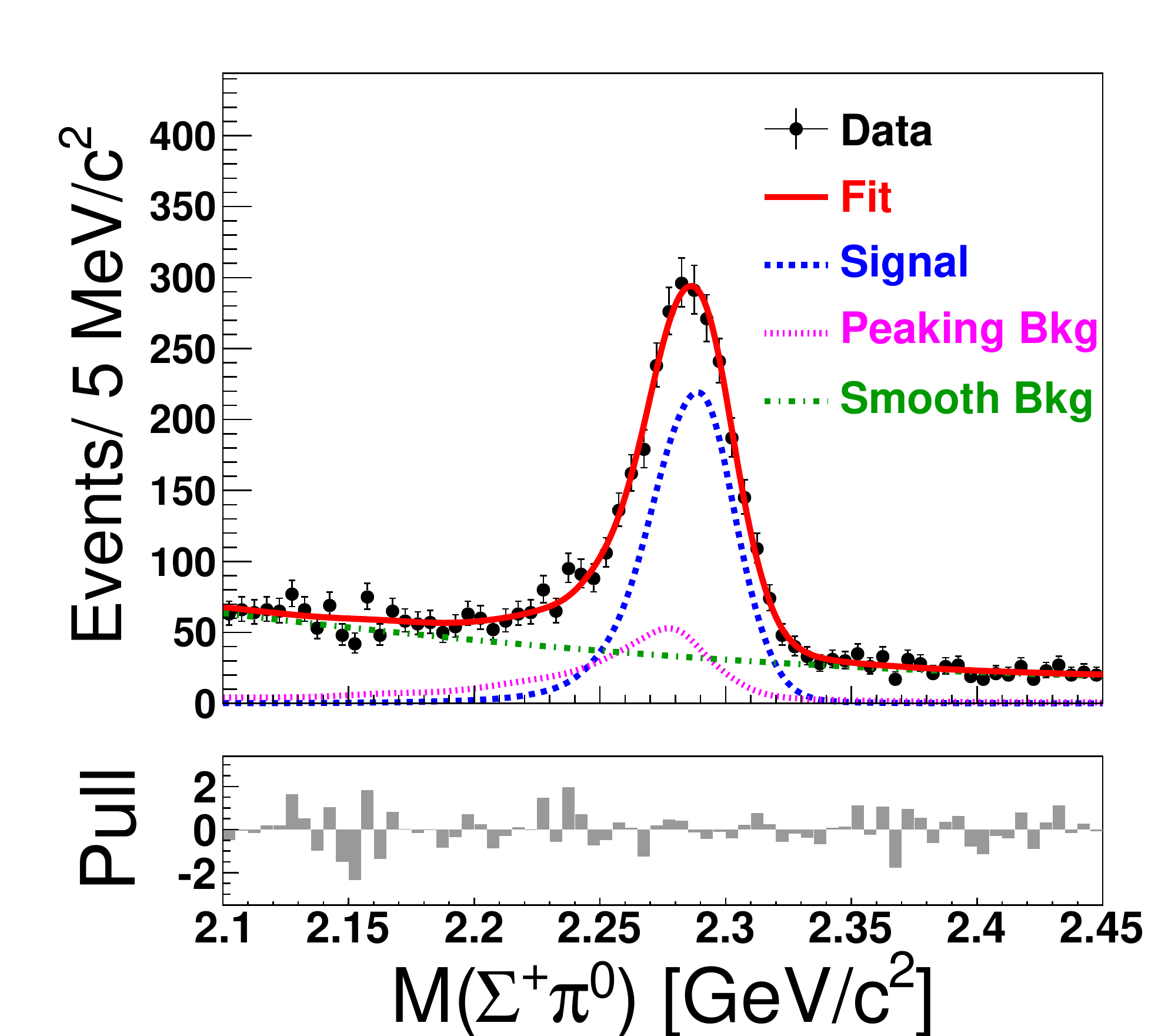}
          \put(-278,115){\large \bf (d)} \put(-117,115){\large \bf (e)}
          \caption{Fits to $M(\Sigma^+\pi^0)$ distributions from data in (a) $-1.0<\cos\theta_{\Sigma^+}<-0.6$, (b) $-0.6<\cos\theta_{\Sigma^+}<-0.2$, (c) $-0.2<\cos\theta_{\Sigma^+}<0.2$, (d) $0.2<\cos\theta_{\Sigma^+}<0.6$, and (e) $0.6<\cos\theta_{\Sigma^+}<1.0$.}
          \label{fig:sgmpi0-sepefit}
     \end{figure*}

\begin{figure*}[htbp]
      \flushleft
          \includegraphics[width=5.5cm]{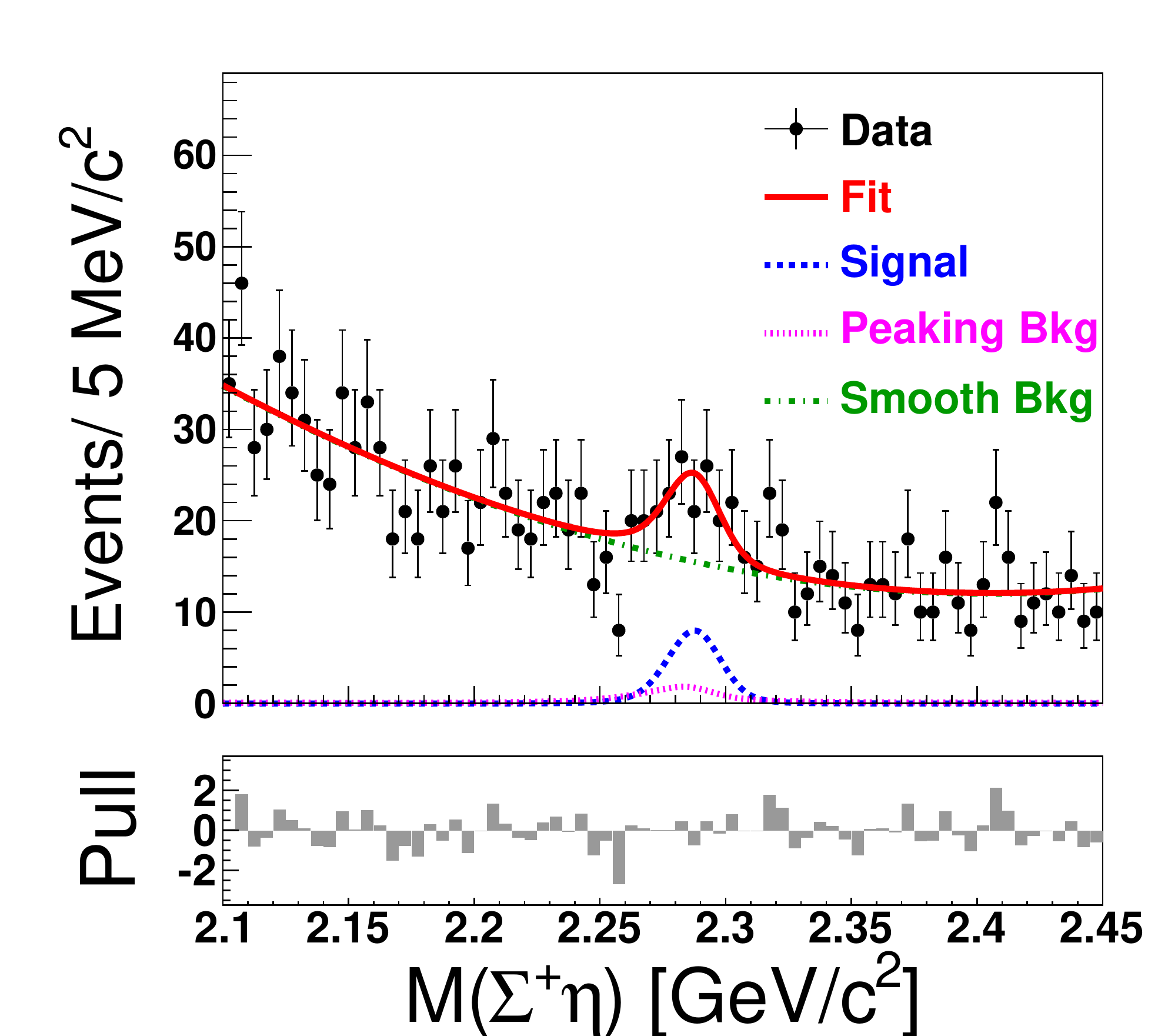}
          \includegraphics[width=5.5cm]{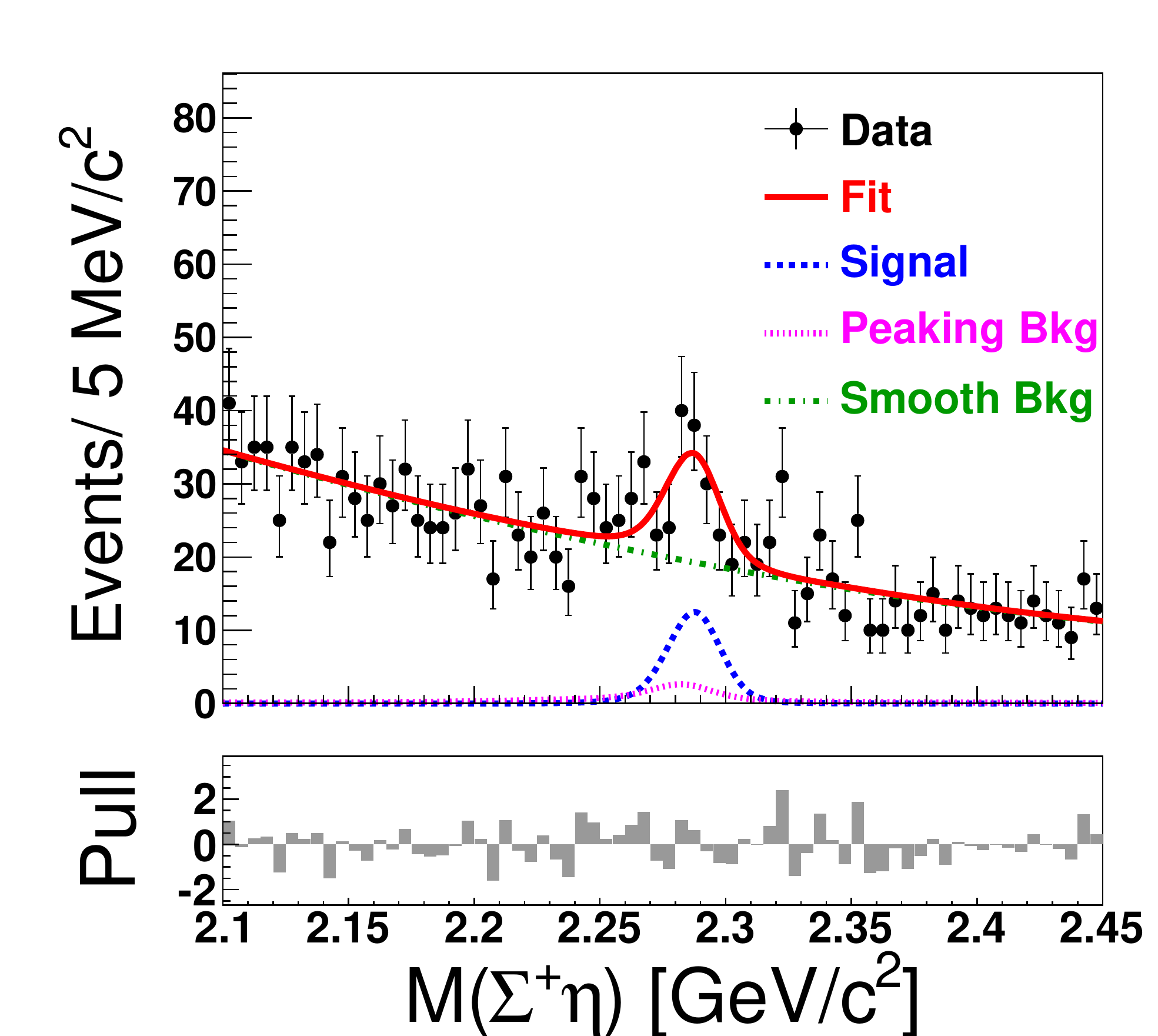}
          \includegraphics[width=5.5cm]{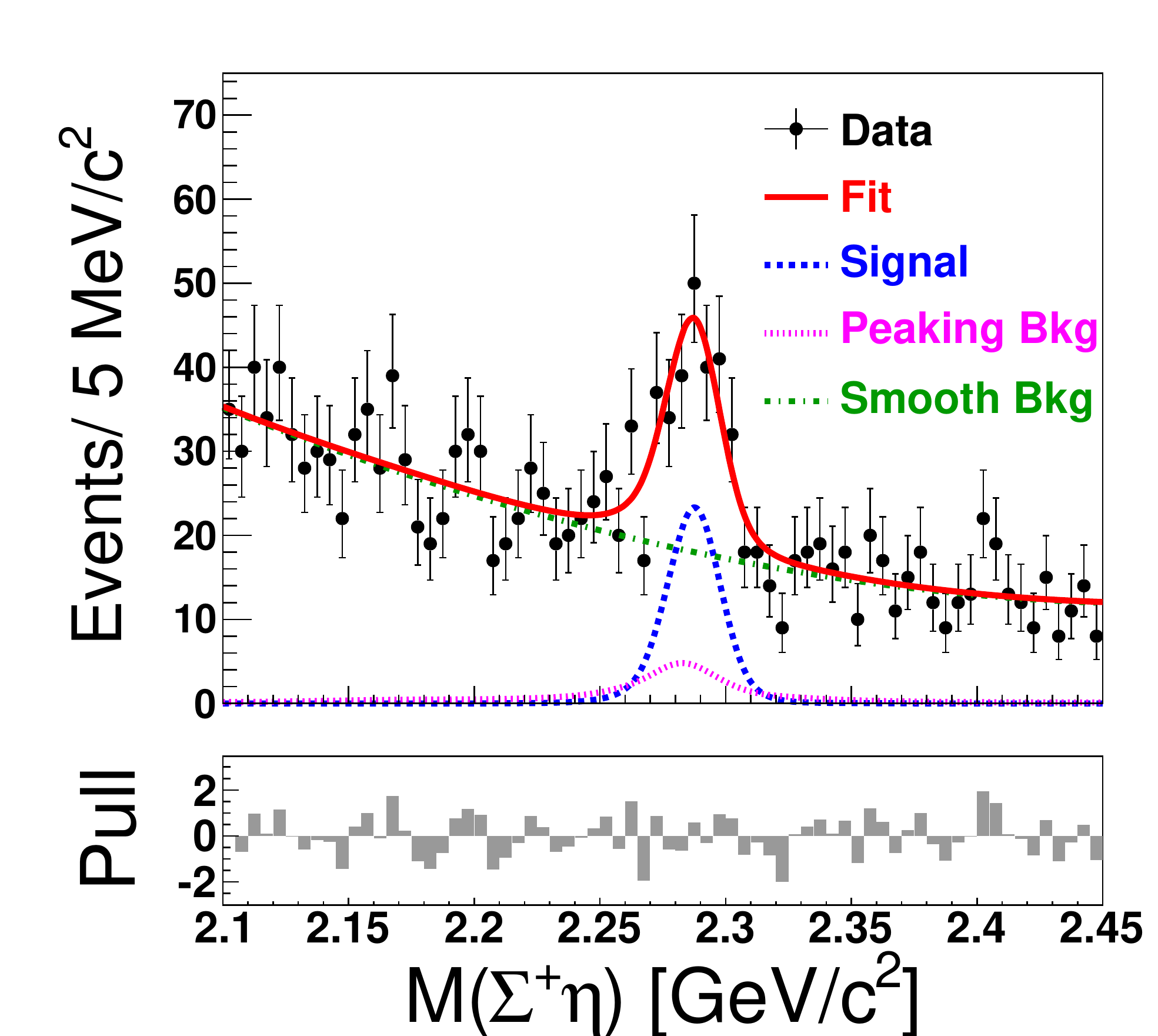}
          \put(-438,115){\large \bf (a)} \put(-277,115){\large \bf (b)}  \put(-115,115){\large \bf (c)}
          \quad
          \includegraphics[width=5.5cm]{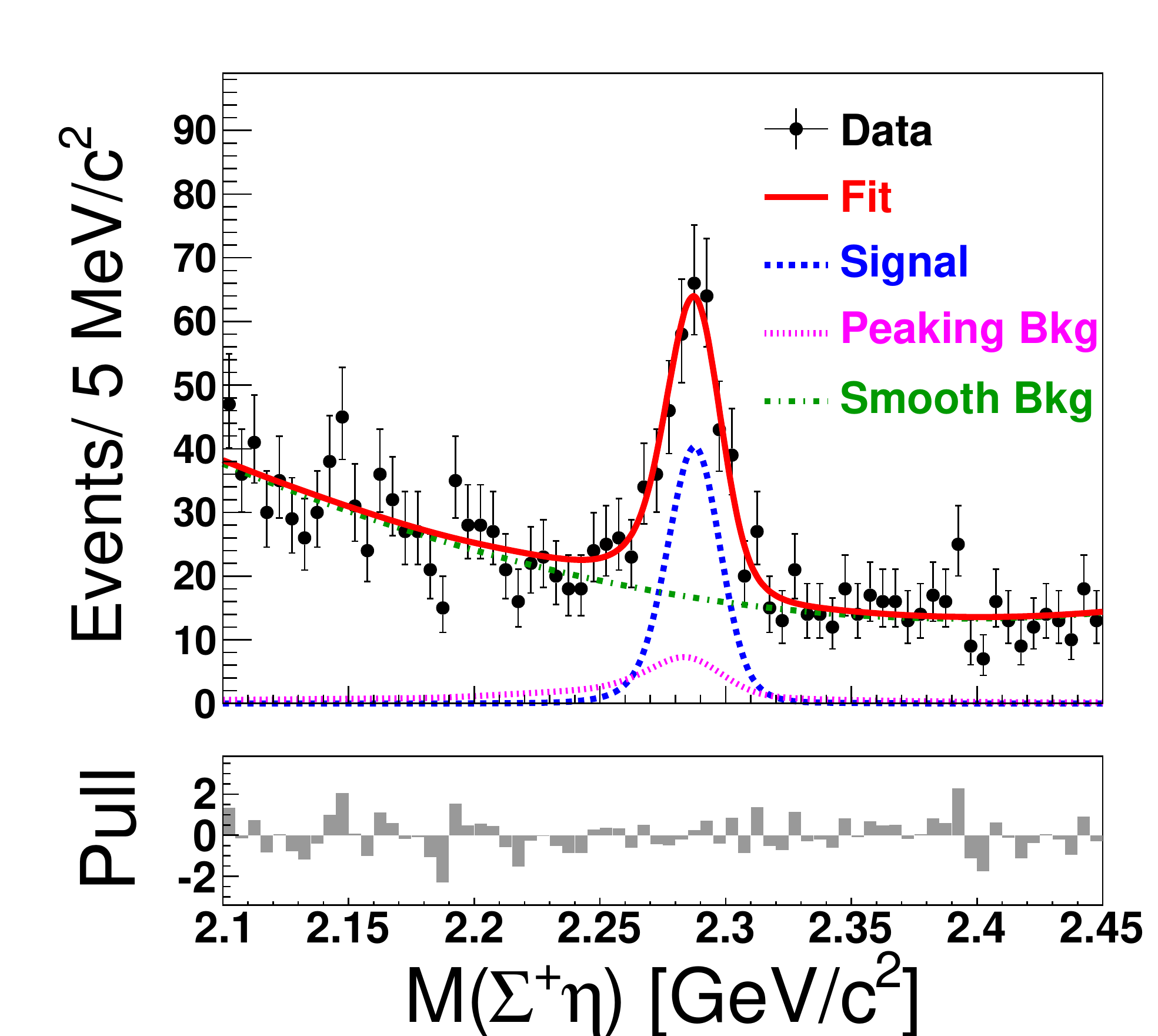}
          \includegraphics[width=5.5cm]{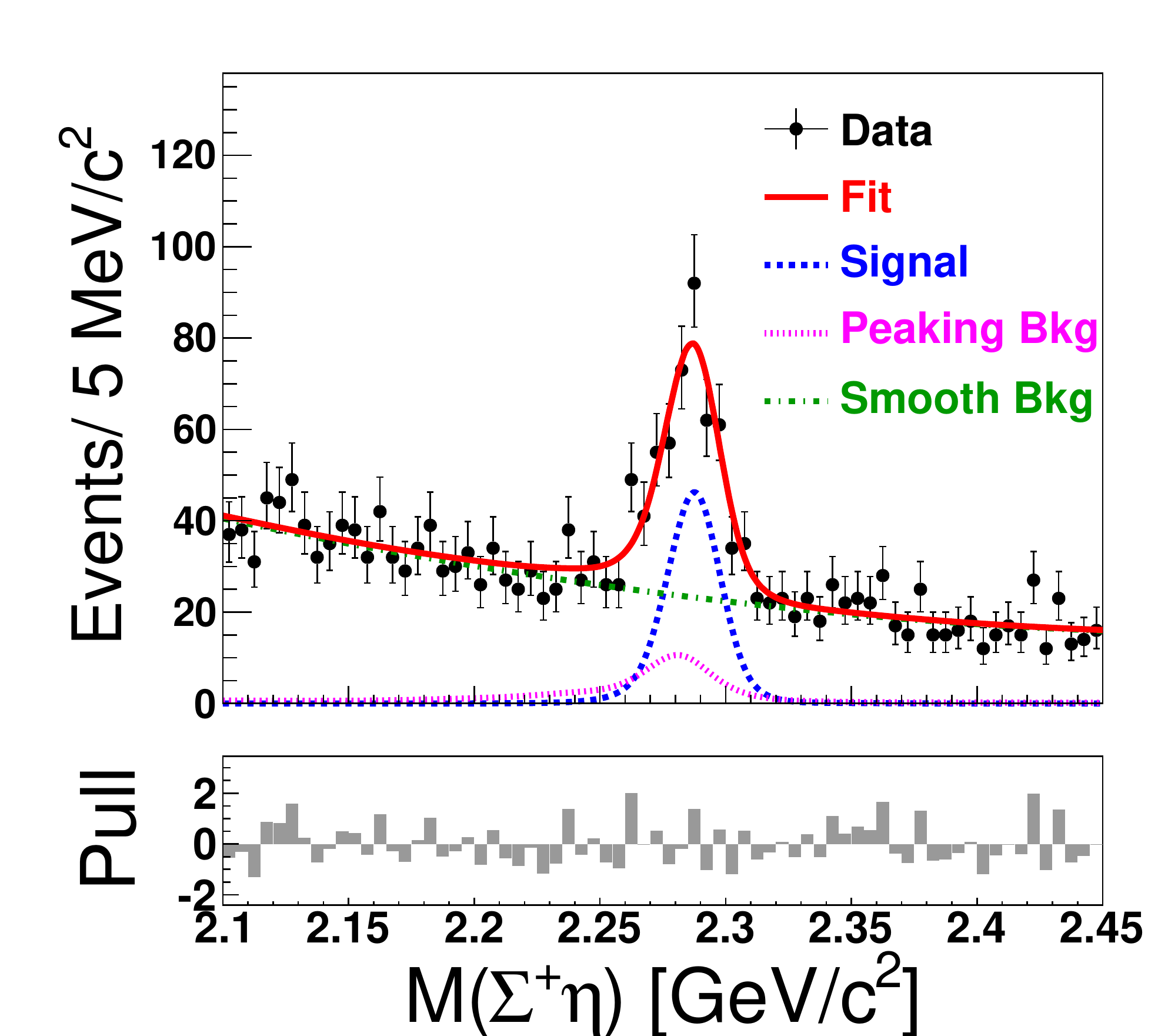}
          \put(-278,115){\large \bf (d)} \put(-117,115){\large \bf (e)}
          \caption{Fits to $M(\Sigma^+\eta)$ distributions from data in (a) $-1.0<\cos\theta_{\Sigma^+}<-0.6$, (b) $-0.6<\cos\theta_{\Sigma^+}<-0.2$, (c) $-0.2<\cos\theta_{\Sigma^+}<0.2$, (d) $0.2<\cos\theta_{\Sigma^+}<0.6$, and (e) $0.6<\cos\theta_{\Sigma^+}<1.0$.}
          \label{fig:sgmeta-sepefit}
\end{figure*}

\begin{figure*}[htbp]
          \flushleft
          \includegraphics[width=5.5cm]{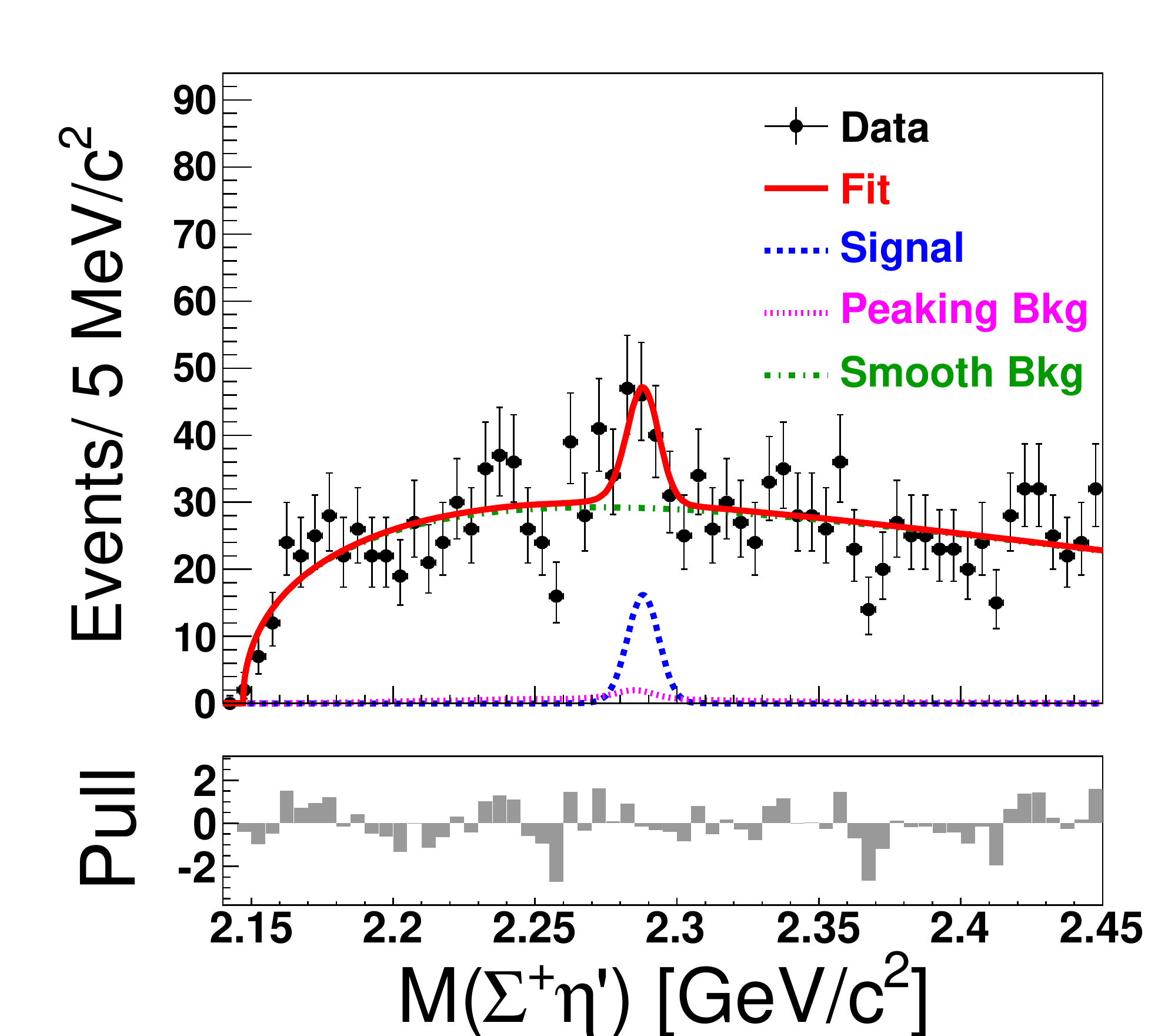}
          \includegraphics[width=5.5cm]{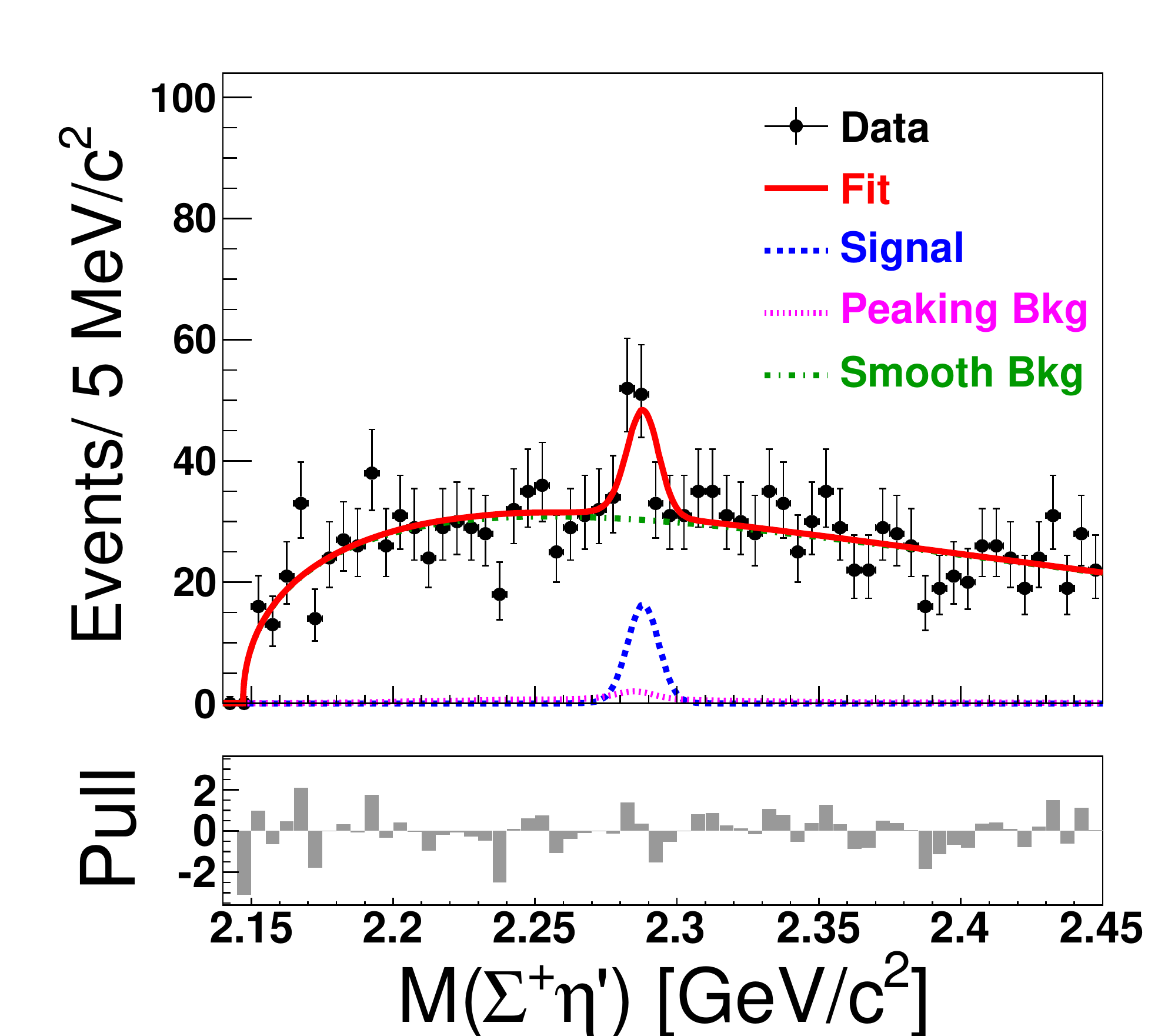}
          \includegraphics[width=5.5cm]{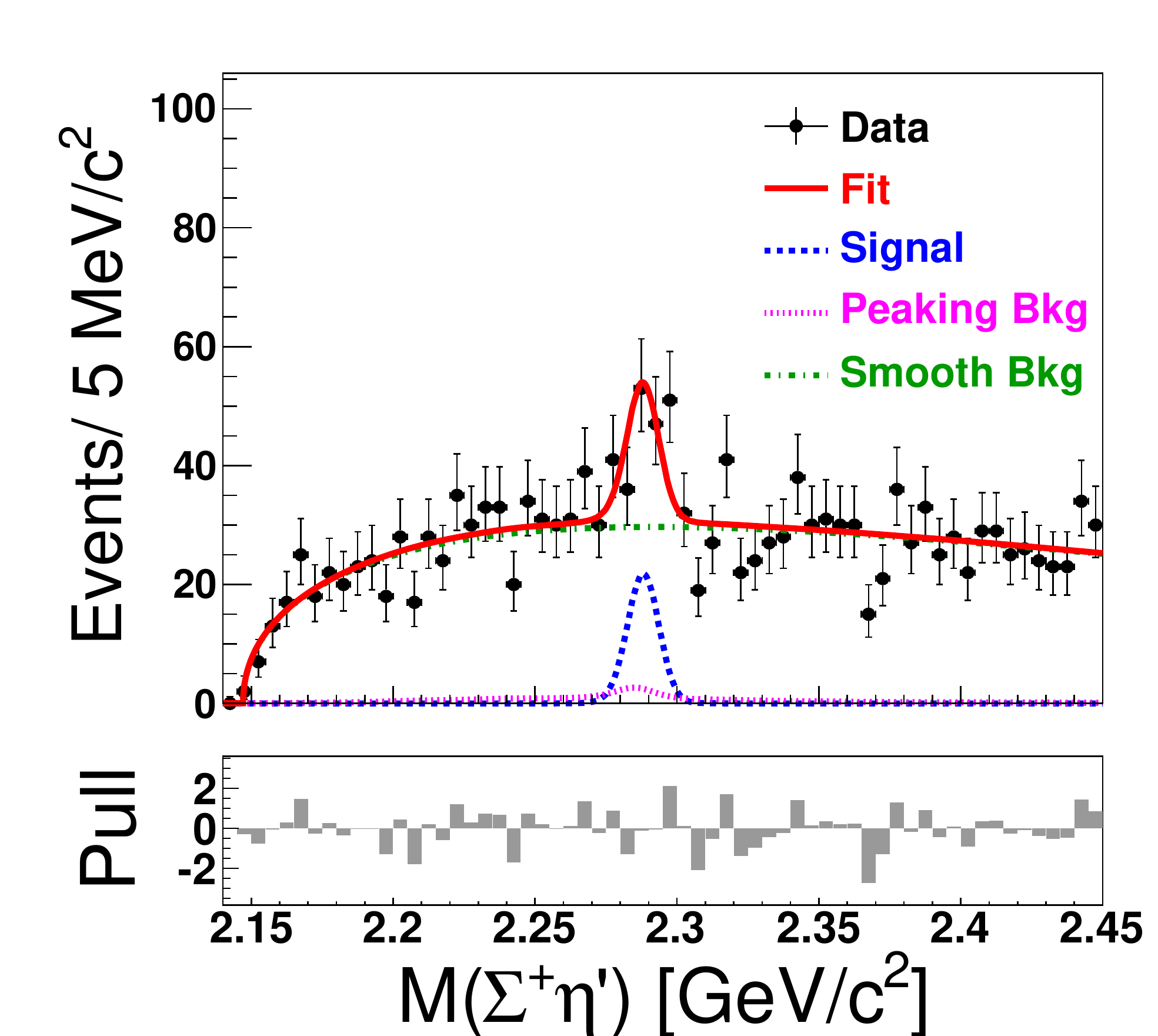}
          \put(-438,115){\large \bf (a)} \put(-277,115){\large \bf (b)}  \put(-115,115){\large \bf (c)}
          \quad
          \includegraphics[width=5.5cm]{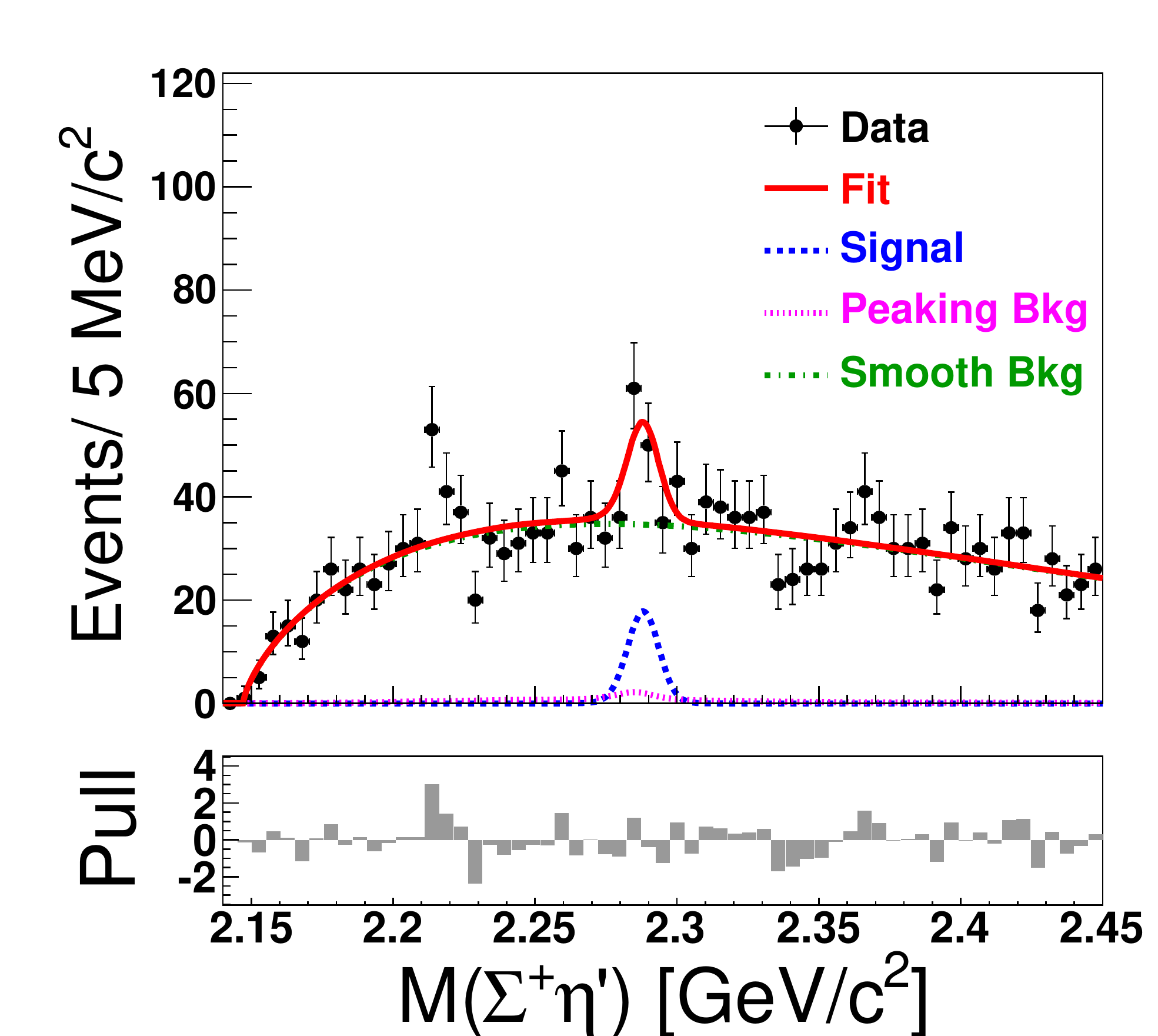}
          \includegraphics[width=5.5cm]{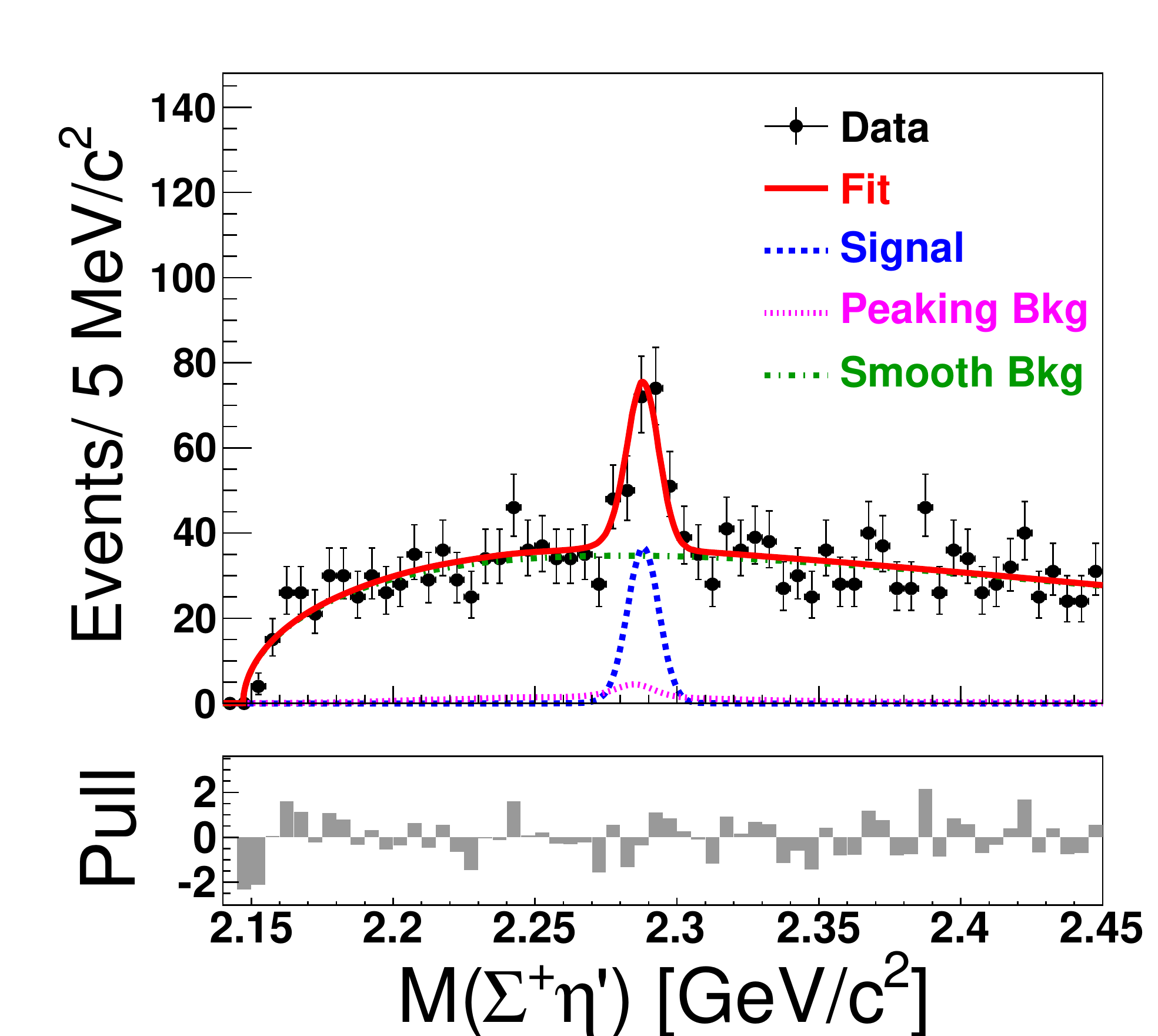}
          \put(-278,115){\large \bf (d)} \put(-117,115){\large \bf (e)}
          \caption{Fits to $M(\Sigma^+\eta')$ distributions from data in (a) $-1.0<\cos\theta_{\Sigma^+}<-0.6$, (b) $-0.6<\cos\theta_{\Sigma^+}<-0.2$, (c) $-0.2<\cos\theta_{\Sigma^+}<0.2$, (d) $0.2<\cos\theta_{\Sigma^+}<0.6$, and (e) $0.6<\cos\theta_{\Sigma^+}<1.0$.}
          \label{fig:sgmetp-sepefit}

\end{figure*}

The fits to the $M(\Sigma^+h^0)$ spectra in bins of $\cos\theta_{\Sigma^+}$ are shown in Fig.~\ref{fig:sgmpi0-sepefit} for $\Sigma^+\pi^0$ mode, in Fig.~\ref{fig:sgmeta-sepefit} for $\Sigma^+\eta$ mode, and in Fig.~\ref{fig:sgmetp-sepefit} for $\Sigma^+\eta'$ mode.

\end{document}